\documentclass[a4paper,fleqn,usenatbib]{mnras}

\usepackage{newtxtext,newtxmath}

\usepackage[T1]{fontenc}
\usepackage{ae,aecompl}
\pdfoutput=1
\usepackage{graphicx}	
\usepackage{amsmath}	
\usepackage{amssymb}	

\usepackage{natbib}
\usepackage{lscape}
\usepackage{longtable}

\newcommand{\FeH}{\rm [Fe/H]}
\newcommand{\ubv}{\ensuremath{UB\,V}}

\newcommand{\logg}{\ensuremath{\log g}}
\newcommand{\Teff}{\ensuremath{T_{\mathrm{eff}}}}

\title[Geneva metallicity calibrations]{Metallicity calibrations for dwarf stars and giants in the Geneva photometric system}

\author[M. Netopil]{
Martin Netopil,$^{1,2}$\thanks{E-mail: mn.netopil@gmail.com}
\\
$^{1}$Department of Theoretical Physics and Astrophysics, Masaryk University, Kotl\'a\v{r}sk\'a 2, 611 37 Brno, Czech Republic\\
$^{2}$Institut f\"ur Astrophysik, Universit\"at Wien, T\"urkenschanzstra{\ss}e 17, A-1180 Wien, Austria\\
}

\date{Accepted 2017 May 2. Received 2017 May 2; in original form 2017 March 6}

\pubyear{2016}

\begin{document}
\label{firstpage}
\pagerange{\pageref{firstpage}--\pageref{lastpage}}
\maketitle

\begin{abstract}
We use the most homogeneous $Geneva$ seven-colour photometric system to derive new metallicity calibrations for early A- to K-type stars that cover both, dwarf stars and giants. The calibrations are based on several spectroscopic data sets that were merged to a common scale, and we applied them to open cluster data to obtain an additional proof of the metallicity scale and accuracy. In total, metallicities of 54 open clusters are presented. The accuracy of the calibrations for single stars is in general below 0.1\,dex, but for the open cluster sample with mean values based on several stars we find a much better precision, a scatter as low as about 0.03\,dex. Furthermore, we combine the new results with another comprehensive photometric data set to present a catalogue of mean metallicities for more than 3000 F and G-type dwarf stars with $\sigma \sim 0.06$\,dex. The list was extended by more than 1200 hotter stars up to about 8500\,K (or spectral type A3) by taking advantage of their almost reddening free characteristic in the new $Geneva$ metallicity calibrations. These two large samples are well suited as primary or secondary calibrators of other data, and we already identified about 20 spectroscopic data sets that show offsets up to about 0.4\,dex.
\end{abstract}




\section{Introduction}

Stellar metallicity is a key parameter that influences stellar evolution, for example, but it is also an important measure for a better understanding of the Milky Way itself. For the latter, open clusters are a valuable tool, because they allow us to accurately determine the age and the distance. The cluster member stars show in general chemical homogeneity, thus an average of stellar data provide us with a very accurate value for this additional parameter as well. However, we only know the metallicity for a limited sample of open clusters. A series of papers studied the current knowledge of open cluster metallicities \citep[][]{paunzen10,heiter14,netopil16}. These works include a compilation of photometric estimates, an evaluation of spectroscopic data, and finally the presentation of 172 objects on a homogeneous metallicity scale, including spectroscopic and photometric estimates. The last reference used the final sample for an investigation of the Galactic metallicity distribution and found among others a first observational evidence of radial migration based on open cluster data.

More efforts are needed to enlarge the sample of open cluster metallicities, but also to improve the accuracy of the data. This holds for photometric estimates, but for spectroscopic measurements as well. Although spectroscopic data can provide us with the highest possible accuracy, the metallicity scale can differ significantly from one study to another \citep[see discussion by][]{heiter14}. Ongoing surveys, such as the $Gaia$-ESO Public Spectroscopic Survey \citep[][]{gilmore12}, and later the data of the $Gaia$ satellite mission itself will certainly contribute to a significant improvement in this respect. 

Despite all these projects, photometric metallicities cannot be considered as outdated, they might still provide valuable results to complement spectroscopic data \citep[][]{netopil16}. From that work and the additional literature, it is evident that one photometric system was not used so far to derive open cluster metallicities:  the $Geneva$ photometric system \citep{golay72}. This seven-colour system allows to derive astrophysical parameters for a broad range of spectral types \citep{kunzli97} and was also used to investigate chemically peculiar stars \citep[e.g.][]{hauck82}, for example. Some empirical metallicity calibrations are available for this system \citep[see e.g.][and references therein]{berthet90}, but these were not systematically applied, neither to open clusters nor to field stars. The last author also found a close correlation between the blanketing parameters in the $Geneva$ system ($\Delta m_2$) and the Str\"omgren-Crawford one ($\delta m_1$). The latter system was already used for some detailed studies of open clusters \citep{anthony-twarog16} or field stars \citep{casagrande11}.

The big advantage of the $Geneva$ system compared to other photometric systems is that it belongs to the most homogeneous ones, the data are based on a unique instrumentation and reduction procedure \citep{cramer99} in both hemispheres. The General Catalogue of Photometric Data \citep[GCPD\footnote{http://gcpd.physics.muni.cz/},][]{mmh97} includes about 44\,000 stars that cover all spectral types and luminosity classes. This high number of stars and numerous new spectroscopic iron abundance measurements makes a revision of the metallicity calibrations worthwhile.

Although this work is mostly motivated to improve the knowledge of open cluster metallicities, it is clear that the limiting magnitude of the $Geneva$ data ($\sim$\,12\,mag) restricts an open cluster sample to nearby objects, which are probably already studied elsewhere. However, a big homogeneous sample allows a profound validation of the derived metallicity scale if compared to other data and possibly to identify  erroneous literature results. As a next step the calibrations can be applied to samples of field stars to verify other data sets, for example.

In this article the term
metallicity is synonymously used with iron abundance \FeH, but we note that photometric approaches in general measure [M/H]. Thus, in the case of non-solar scaled abundances, the calibrated \FeH\ value might deviate from the actual iron abundance. The article is arranged as follows: in Sect. \ref{specmetal} we discuss the compilation of spectroscopic metallicities, Sect. \ref{metalcal} presents calibrations for dwarf stars and giants, Sect. \ref{addcomp} shows additional comparisons of the metallicity scale and presents open cluster metallicities. In Sect. \ref{sect:photsamples} we make use of the new calibrations and present catalogues of photometric metallicities. Finally, Sect. \ref{sect:conclusion} summarises the work.

\section{Spectroscopic metallicity scale}
\label{specmetal}
The calibration of the $Geneva$ blanketing parameters in terms of \FeH\ needs a sufficiently large sample of stars with well-determined astrophysical parameters. We therefore use literature results listed in the PASTEL catalogue of stellar parameters \citep[][version 2016-03-18]{soubiran16}, but exclude works that are published before 1990; the restriction adopted by \citet{heiter14}. We checked the references in the database if they provide independent results, to avoid a bias owing to identical entries if an author or author group apparently adopt their previous results for a study in another context. This was noticed for example in the works by \citet{mishenina08,mishenina12}, \citet{trevisan11} and \citet{trevisan14}, or \citet{luck05,luck06}. Furthermore, we exclude the results by \citet{sousa08}, because the data were re-analysed by \citet{tsantaki13} using the same automatic method but with an optimised iron line list. The differences in \FeH\ are very small as shown by the last reference. Thus, we will introduce a bias if both results are adopted, in particular, because the data set is one of the largest for cool dwarfs. We also corrected some minor inconsistencies in the database (e.g. identical double entries from the same reference) and informed the maintainer about it. 

The use of a single homogeneous data set does not provide the needed number of stars over a broad range of temperature, luminosity, and metallicity. On the other hand, merging different data sets might introduce distortions and a larger scatter owing to different scales. The overlap between individual spectroscopic data sets is often too small to conclude about possible offsets, but we can use the photometric metallicities in an iterative way to identify stronger deviating spectroscopic scales. We start with mean (uncorrected) spectroscopic metallicities to derive initial metallicity calibrations as outlined in the following section and investigate differences between the photometric scale and individual spectroscopic data sets. The last were corrected by the identified offsets and the procedure was repeated several times until we do not notice any additional deviation. A comparison of the mean uncorrected metallicities and the final corrected ones shows that the standard deviation has improved by about 20 per cent.

Most spectroscopic datasets are well scaled, but for some we are able to conclude an offset as listed Table \ref{tab:offsets}. The study by \citet{roederer14} shows by far the largest offset, which is in good agreement with the differences to other data presented by the reference itself (their table 17). The works by \citet{varenne99} and \citet{gebran10} analysed hotter stars in the Hyades open cluster, but obtained a too low metallicity compared with the currently accepted value of [Fe/H] = 0.13 $\pm$ 0.05\,dex based on high quality spectroscopic data and [Fe/H] = 0.15 $\pm$ 0.03\,dex using results from three photometric metallicity calibrations \citep{heiter14,netopil16}. The largest spectroscopic data sets for dwarfs with more than 300 stars in common \citep{bensby14,ramirez13,sousa11,tsantaki13,valenti05} agree at a $\pm 0.01$\,dex level with $\sigma \sim 0.11$\,dex, which is confirmed by the direct comparisons as shown for example by \citet{bensby14}.

The largest samples for giant stars are by \citet{hekker07} and \citet{mcwilliam90}. We derive an offset for the last reference, whereas the other one agrees with the final photometric scale ($0.01 \pm 0.10$\,dex, 186 stars). The offset is also confirmed by a direct comparison of the two spectroscopic data sets.

We also need to verify that the metallicity scales for giants
and dwarfs agree. Unfortunately, only a few studies provide a high number of giants and dwarfs that overlap with our photometric sample. However, based on the data by \citet{dasilva11,dasilva15}, \citet{jofre15}, and \citet{randich99} we conclude that the metallicity scales agree better than 0.03\,dex.

\begin{table}
\caption{Derived offsets of deviating metallicity scales. The differences are calculated as photometric $-$ spectroscopic scale. } 
\label{tab:offsets} 
\centering 
\begin{tabular}{l l l } 
\hline 
Reference & $\Delta$\FeH & Stars   \\ 
\hline 
\citet{bond06} & +0.09 $\pm$ 0.07 & 94 \\
\citet{brewer06} & +0.11 $\pm$ 0.08 & 12 \\
\citet{clementini99} & $-$0.09 $\pm$ 0.10 & 38 \\
\citet{dasilva06} & $-$0.08 $\pm$ 0.09 & 37 \\
\citet{fulbright00} & +0.12 $\pm$ 0.11 & 54 \\
\citet{gebran10} & +0.08 $\pm$ 0.05 & 23 \\
\citet{jones92}  & $-$0.10 $\pm$ 0.07 & 9 \\
\citet{jones11}  & $-$0.08 $\pm$ 0.07 & 73 \\
\citet{luck14} & $-$0.16 $\pm$ 0.10 & 22 \\
\citet{mallik98} & $-$0.11 $\pm$ 0.09 & 13 \\
\citet{mcwilliam90} & +0.10 $\pm$ 0.08 & 280 \\
\citet{morel14} & $-$0.08 $\pm$ 0.08 & 9 \\
\citet{pasquini94} & +0.13 $\pm$ 0.09 & 24 \\
\citet{paulson06} & $-$0.14 $\pm$ 0.14 & 9 \\
\citet{pompeia11} & $-$0.12 $\pm$ 0.10 & 8 \\
\citet{qui02} & $-$0.15 $\pm$ 0.10 & 10 \\
\citet{roederer14} & +0.38 $\pm$ 0.10 & 10 \\
\citet{thevenin99} & $-$0.17 $\pm$ 0.12 & 62 \\
\citet{tomkin92} & +0.21 $\pm$ 0.13 & 17 \\
\citet{varenne99} & +0.21 $\pm$ 0.06 & 15 \\
\citet{zhao90} & +0.24 $\pm$ 0.15 & 12 \\
\citet{zhao01} & $-$0.12 $\pm$ 0.08 & 28 \\
\hline 
\end{tabular}

\end{table}

Finally, we derived mean values for \Teff, \logg, and \FeH\ if multiple determinations are available for an object. We note that we do not make comparisons of the gravity and temperature scales, because these are not that critical for our purpose and are in general only used as additional guidance for the sample selection.

\section{Metallicity calibration in the Geneva system}
\label{metalcal}
\subsection{Main sequence stars}
\label{dwarf_cal}
For most photometric systems that allow an estimate of metallicity, a standard sequence based on Hyades stars is used to correct for the colour (temperature) dependency and to derive the respective blanketing parameters \citep[see e.g.][for the \ubv\ and $uvbyH\beta$ systems]{crawford75}. To our knowledge, the last update of the Hyades sequence in the $Geneva$ photometric system is given by \citet{hauck91}, which was used by \citet{berthet90} to derive an updated metallicity calibration in this system. Many more stars were observed in the Geneva system, but also the number of stars with known iron abundance increased significantly after this work. Thus, the data allow a verification and improvement of both, the standard sequence and the metallicity calibration.

\begin{figure}
\centering
\resizebox{\hsize}{!}{\includegraphics{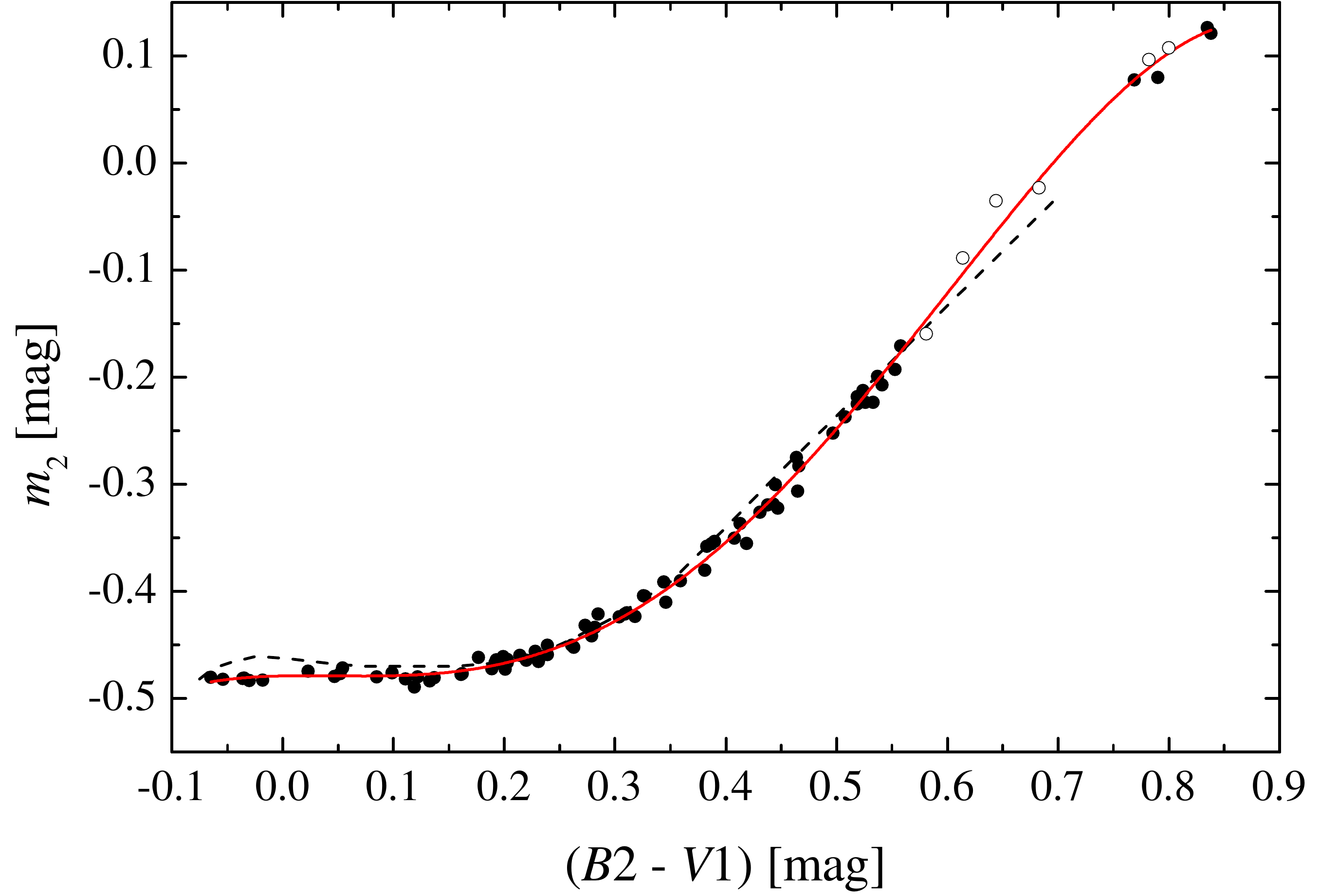}}
\caption{Reference sequence in the $(B2-V1)/m_2$ plane based on stars in the Hyades open cluster with a high photometric weight (filled circles) and the fitted fourth order polynomial as given in Table \ref{tab:sequence} (red line). Open circles represent additional cooler type stars down to about spectral type K5 with lower photometric weights. The black dashed line is the standard sequence given by \citet{hauck91}. }
\label{fig:hyades}
\end{figure}

To construct the reference sequence we queried for $Geneva$ photometry of Hyades stars in the GCPD. The data include a weight related to the photometric quality, and we only adopt stars that have a weight of at least three and photometric errors less than 0.01\,mag. These objects can be considered as the most often observed and constant stars. Furthermore, we excluded chemically peculiar stars or objects that are listed as binary components. In addition to the photometric data, available kinematic membership data \citep{perryman98} were used to extract the most probable cluster members. The colour range $-0.065 \le (B2-V1) \le 0.55$\,mag, corresponding to spectral types from about A3 to K1, is very well covered, but our quality criteria lower the number of cooler type objects. Thus, we also included $Geneva$ data of some objects with a lower photometric weight, resulting in 86 stars in total. The final Hyades sequence in the $(B2-V1)/m_2$ plane is shown in Fig. \ref{fig:hyades}, and we recall that the $Geneva$ blanketing index is defined as $m_2 = (B1 - B2) - 0.457 (B2 - V1)$. The colour index $(B2-V1)$ shows a good temperature sensitivity over a wide colour range and was therefore used also in the previous metallicity calibrations. A fourth order polynomial is needed to represent the data best and we list the derived coefficients in Table \ref{tab:sequence}.  The final colour corrected blanketing parameter is then defined as $\Delta m_2 = m_2{\rm (obs)} - m_2{\rm (std)}$.

Figure \ref{fig:hyades} shows that the standard sequence by \citet{hauck91} does not match the observations well. Differences up to about 0.02\,mag are recognised for mid-A and G-type stars. The reason for the discrepancy is unknown, but might be related to  the low number of stars used by \citet{hauck91}. They mentioned the use of 21 objects with $(B2-V1) < 0.34$\,mag, whereas we compiled a twice larger number of stars in this colour range. However, we will later discuss the influence of this difference on the results.

\begin{table}
\caption{Standard sequences for dwarfs and giants. 
 } 
\label{tab:sequence} 
\begin{center} 
\begin{tabular}{l c c c } 
\hline 
 & $m_2$ &  $U_t$ & $U_t$(corr) \\ 
\hline 
$a_0$ & $-$0.479 (2) & +1.45 (12) & +1.94     \\
$a_1$ & +0.02   (3)   & $-$3.11 (134) & $-$8.91   \\
$a_2$ & $-$0.65 (22) & +21.34 (503) & +44.66  \\
$a_3$ & +4.97   (47) & $-$31.38 (772) & $-$67.20  \\
$a_4$ & $-$3.82 (31) & +14.45 (415) & +33.17  \\
N     &  86                &  40 & \\
\hline
$a_0$     &               & +0.60 (6) & 0.89 \\
$a_1$     &               & +3.96 (21) & 2.80  \\
$a_2$     &               & $-$1.19 (19) &  \\
N     &                  &  131 & \\

\hline 
\end{tabular}
\end{center} 
\medskip
Notes: The upper panel lists the results for dwarfs and the lower panel for the giant stars. The regressions are given in the form a$_0$ + a$_1$ $X$ + a$_2$ $X^2$ ... with $X$ representing $(B2-V1)$ and $t$ for $m_2$ and $U_t$, respectively. 
The number of used stars (N) are listed in the last lines. The errors of the last given digits are listed in parentheses. The valid ranges are $-0.065 \leq (B2-V1) \leq 0.84$\,mag and $0.17 \leq t \leq 0.78$ for dwarfs and $0.28 \leq t \leq 0.90$ for giants, respectively.

\end{table}

\begin{table*}
\caption{Metallicity calibration of main sequence stars.  } 
\label{tab:calib} 
\begin{center} 
\begin{tabular}{l c c c c c} 
\hline
 & \FeH$_{\Delta m_2}$   & \FeH$_{\Delta m_2}$  & \FeH$_{\Delta m_2}$   & \FeH$_{\Delta U_t}$ &  \FeH$_{\Delta U_t}$ \\ 
 &   $(B2-V1) < 0.21$         & $-0.065 \leq (B2-V1) \leq 0.40$	& $(B2-V1) > 0.35$        & $-0.65 < \Delta U_t < -0.07$ & $-0.07 \leq \Delta U_t < 0.20$ \\
\hline 
$a_0$ &   +0.13            &   +0.13 (0.01)   &  0.13             &  +0.29 (0.06) & +0.13 \\
$a_1$ &  +8.12  (0.70)     &  +8.61  (0.67)  &  5.66 (0.34)      & +4.35 (0.14) & +2.11 (0.49)\\
$a_2$ &  $-$105.16 (10.98) &  $-$103.23 (11.29) &  $-$35.49 (3.83)  &             & \\
$a_3$ &                    &  $-$19.62 (5.61) &                   &             & \\
$a_4$ &                    &  +190.80 (35.83) &                   &             & \\
\hline
N          &  70             &   501          &   166          & 197   & 155\\
$\sigma$[dex]      & 0.10    &  0.09        &   0.08                   &  0.10  & 0.06\\

\hline 
\end{tabular}
\end{center}
\flushleft
\medskip
Notes: The regressions are given in the form a$_0$ + a$_1$ $X$ + a$_2$ $X^2$ + a$_3$ $XY^2$ + a$_4$ $X^2Y$, with $X$ representing $\Delta m_2$ or $\Delta U_t$, respectively, and $Y = (B2-V1)$. The errors are given in parentheses. The number of calibration stars in the respective colour range and the standard deviation of the metallicity calibrations are listed in the last lines.
\end{table*}

We matched the $Geneva$ catalogue with the spectroscopic data presented in Sect. \ref{specmetal} and we adopt only stars closer than 150\,pc using the Hipparcos parallax \citep{leeuwen07} to keep the influence of interstellar reddening as small as possible. Furthermore, we exclude variable stars and binary components using the flags in the $Geneva$ and Hipparcos catalogues. The latter criterion helps to minimise errors owing to possible misidentifications. However, we also checked the visual magnitudes in the two catalogues and removed about 60 stars that show differences larger than 0.3\,mag. Chemically peculiar objects were excluded using the catalogue for Am and Ap stars by \citet{renson09} and the recent $\lambda$ Boo stars catalogue by \citet{murphy15}.

The cleaned starting sample includes about 2900 stars in the colour range of the reference sequence ($-0.065 \le (B2-V1) \la 0.80$\,mag). From this sample, we selected main sequence objects based on the compiled surface gravity (\logg\ $\ga$ 3.8) and the spectral types listed by \citet{skiff14}. Stars with at least three spectroscopic metallicity determinations are in general used as the calibration sample to minimise the influence of single erroneous results, but we adopted all objects with $(B2-V1) \leq 0.20$\,mag to increase the number of hotter stars in the sample. We noticed that $\Delta m_2$ becomes unreliable for stars later than about spectral type G2, thus we set the cool colour limit to $(B2-V1) \leq 0.40$\,mag. The final sample includes 527 stars, with a median of five iron abundance measurements per star and a median $\sigma$ of 0.05\,dex. 

\begin{figure}
\centering
\resizebox{\hsize}{!}{\includegraphics{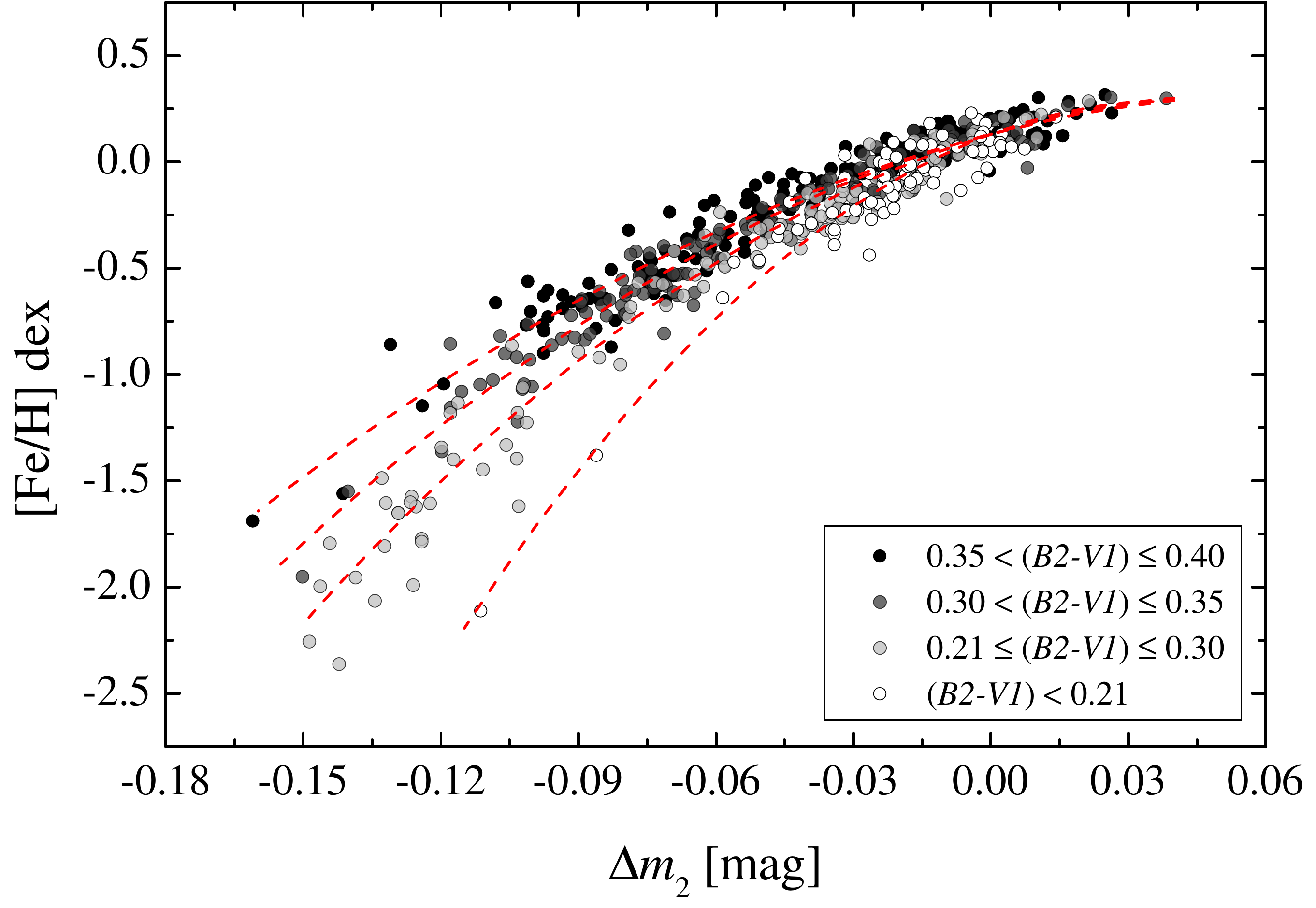}}
\caption{Calibration sample of main sequence stars in the $\Delta m_2$/[Fe/H] plane, grouped by colour ranges. The dashed lines represent polynomial fits to the respective colour ranges.}
\label{fig:calibration}
\end{figure}

\begin{figure}
\centering
\resizebox{\hsize}{!}{\includegraphics{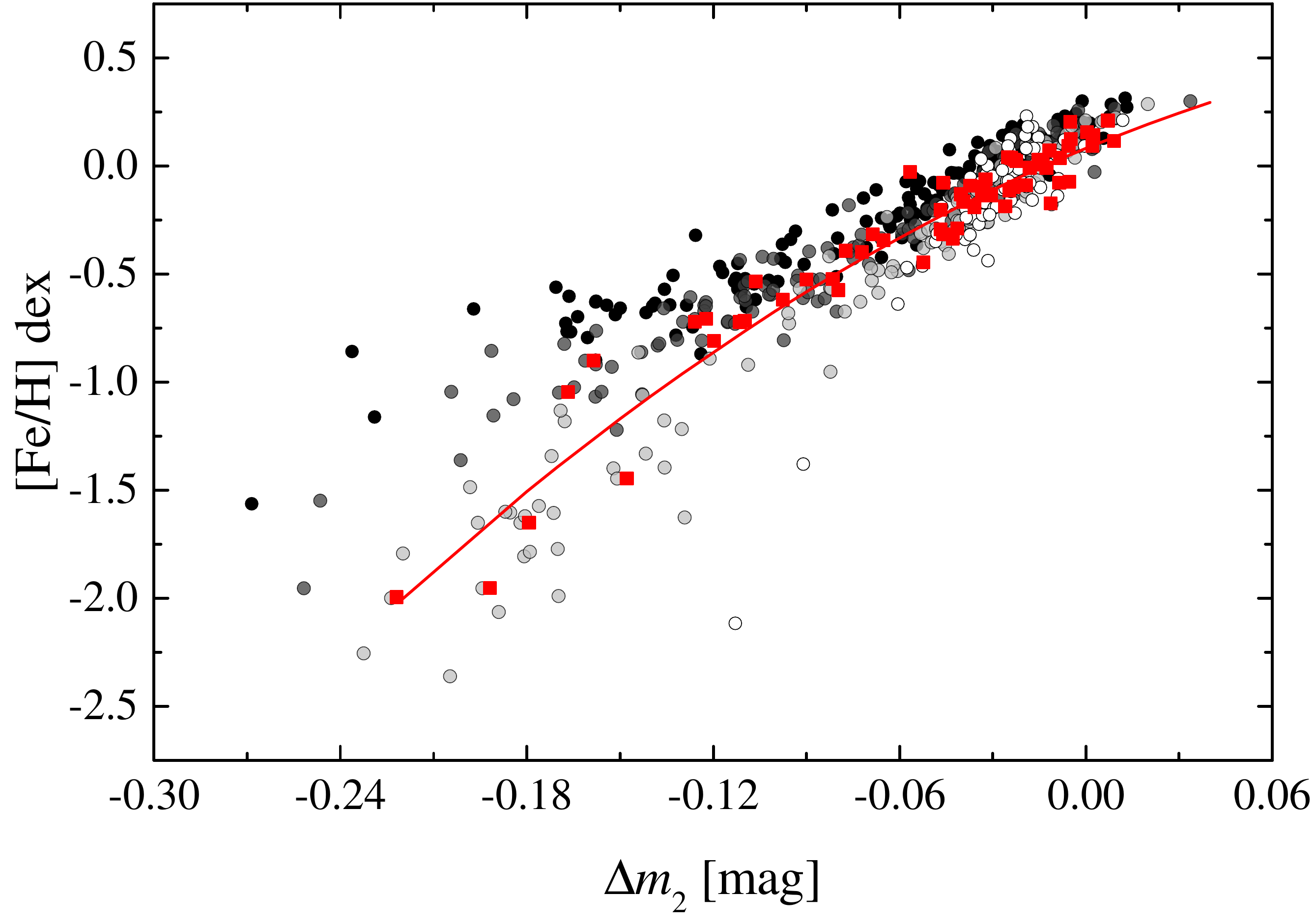}}
\caption{Sample and symbols as in Fig. \ref{fig:calibration}, but calibrated as by  \citet{berthet90}. Red squares represent 63 stars in common with this author, and the red line shows his second order polynomial fit.}
\label{fig:berthet}
\end{figure}

Figure \ref{fig:calibration} shows the calibration sample in the $\Delta m_2$/[Fe/H] plane. The scatter increases towards underabundant objects, but which is obviously caused by the influence of blanketing on the colours. We are faced with the fact that a given $\Delta m_2$ value does not represent the same metallicity at different colours, we therefore apply a multivariate regression with the variables $(B2-V1)$ and $\Delta m_2$ as listed in Table \ref{tab:calib}.
We exclude four apparent outliers for the final calibration, and 22 Hyades stars to not include this cluster twice in the calibration procedure and to test the zero points of the fits. However, the derived zero point of 0.13 $\pm$ 0.01\,dex is in excellent agreement with literature values for the Hyades. Table \ref{tab:calib} also lists polynomial fits to the hottest and coolest objects. These should be applied to underabundant stars ($\Delta m_2 < -0.05 / -0.07$ for the hottest/coolest range) to achieve a better accuracy. The derived zero points are 0.11 $\pm$ 0.02\,dex and 0.14 $\pm$ 0.01\,dex for the hot and cool sample, respectively, thus we fix them to the spectroscopic metallicity value of the Hyades and recalculated the regressions. 

Although not explicitly mentioned, \citet{berthet90} adopted the empirical correction $\Delta (B2-V1) = 1.2(\Delta m_2 + 0.06)$ by \citet{hauck73} to take the influence of blanketing on the colours into account, and lists already the corrected colour index for the stars. This correction was applied to underabundant objects ($\Delta m_2 \le -0.06$\,mag) that are cooler than $(B2-V1) = 0.23$\,mag. However, this procedure results in an overcorrection of cool metal-weak stars and excludes hot objects. Figure \ref{fig:berthet} shows the sample of Fig. \ref{fig:calibration}, but calibrated similarly as \citet{berthet90}, adopting the standard sequence by \citet{hauck91} as well. We note that \citet{berthet90} compiled 164 stars with $(B2-V1) \lesssim 0.4$\,mag, but most are hotter objects. Owing to this bias and the number of underabundant objects (only seven stars with \FeH\ $<-1.0$\,dex), the author has not noticed the inappropriateness of the applied correction. Only 63 stars of the list by \citet{berthet90} are included in our calibration sample owing to the selection criteria, but the adopted iron abundances are in reasonable agreement. Furthermore, the use of the standard sequence by \citet{hauck91} results in an additional spread and colour dependency of the photometric metallicities, the mean metallicity of cool and hot stars at  $\Delta m_2 = 0$\,mag differs by almost 0.10\,dex. 

\begin{figure}
\centering
\resizebox{\hsize}{!}{\includegraphics{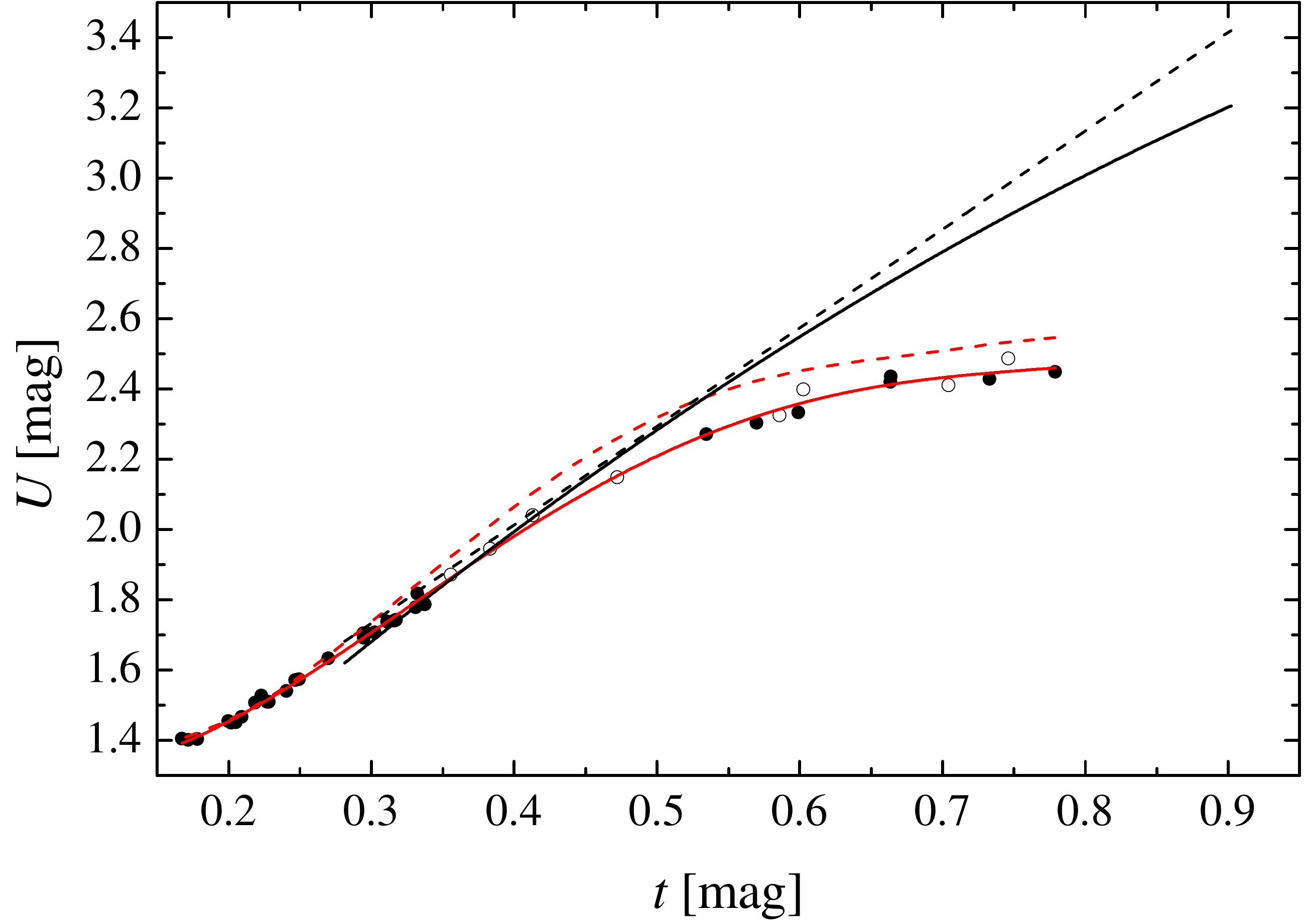}}
\caption{Reference sequences in the $t / U$ plane. Symbols are the same as in Fig. \ref{fig:hyades}. The solid red line is the polynomial fit to the observed sequence of dwarfs in the Hyades, and the red dashed line the corrected sequence as discussed in the text. As comparison, the black solid and dashed lines show the observed and corrected sequences of giant stars.}
\label{fig:hyades_gu}
\end{figure}

Our new metallicity calibration in terms of $\Delta m_2$ provides an accuracy of about 0.10\,dex over the investigated colour range. The calibration derived for the coolest stars can be used for objects as cool as $(B2-V1) = 0.45$\,mag ($\sim$ G5), but a larger error of 0.15\,dex has to be accepted. For cooler stars, $\Delta m_2$ seems not usable anymore, but for stars of spectral type G and later, the $Geneva$ $U$ colour is well suited as blanketing indicator \citep{grenon78}. However, instead of $(B2-V1)$ we use the parameter $t = (B2-G)-0.39(B1-B2)$ \citep{grenon81} to define the reference sequence. For cool stars, the parameter is little influenced by metallicity, thus it is a good temperature indicator \citep{melendez03}. The sequence of the Hyades in the $t / U$ plane in the interval $0.17 < t < 0.78$\,mag, corresponding to early G to late K-type stars, is shown in Fig. \ref{fig:hyades_gu} and the coefficients of the fit are listed in Table \ref{tab:sequence}. The blanketing parameter $\Delta U_t = U(obs) - U_{t}(std)$ shows a colour dependency, comparable to the ultraviolet excess $\delta (U-B)$ in the Johnson \ubv\ system. Therefore, guillotine factors were introduced to normalize $\delta (U-B)$ to the value at $(B-V)=0.6$\,mag \citep[see discussions by][]{karatas06,karaali11}.

The early G-type stars in the calibration sample, which was compiled in the same way as for $\Delta m_2$, were used as a reference and we derived corrections for $\Delta U_t$ as a function of $t$. To provide a convenient application, we calculated a corrected standard sequence, listed as $U_{t}$(corr) in Table \ref{tab:sequence}, which should be used instead of the observed sequence in the Hyades. Owing to this procedure, it is difficult to provide a direct error estimate of the corrected sequence, but based on the scatter of the data we expect that the errors are about three times larger than for the observed Hyades sequence. We can now adopt the complete sample of 370 stars in the corresponding colour range to derive the final metallicity calibration. A two-step linear regression appears as the most reliable relation, and we derive a zero point of \FeH\ = +0.12 $\pm$ 0.01\,dex for $\Delta U_t$  after excluding 3$\sigma$ outliers and stars in the Hyades. We again fix the zero point to the spectroscopic mean value of the Hyades and derived the final coefficients of a piecewise linear function as listed in Table \ref{tab:calib}. The change in the distribution occurs at solar metallicity ($\Delta U_t = -0.07$\,mag or [Fe/H]=$-$0.02\,dex).

\begin{figure}
\centering
\resizebox{\hsize}{!}{\includegraphics{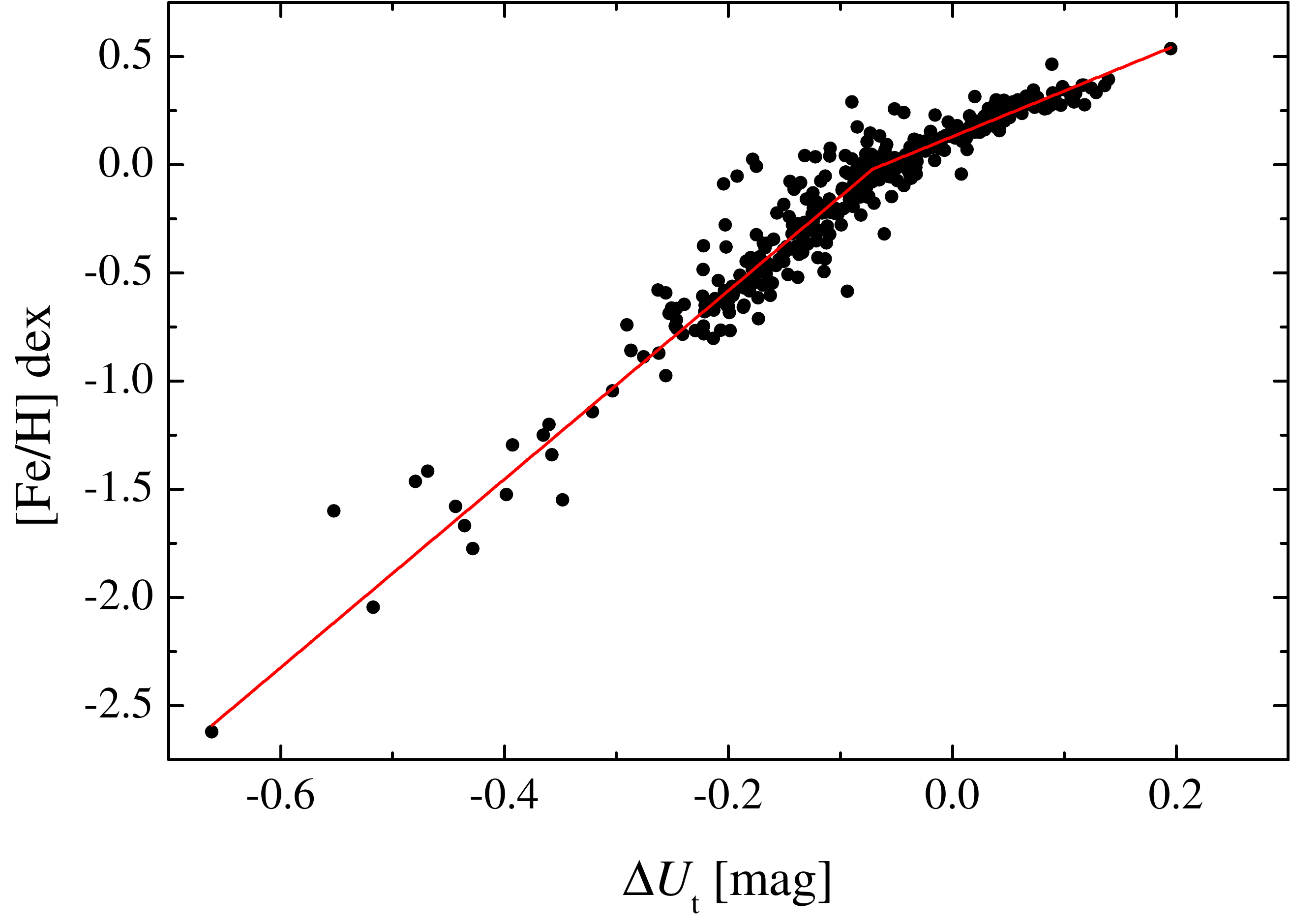}}
\caption{Calibration sample of 444 main sequence stars in the $\Delta U_t$/[Fe/H] plane. The line is the fit of a piecewise linear function as listed in Table \ref{tab:sequence}.}
\label{fig:calibration_dmU}
\end{figure}

The knowledge of interstellar extinction is important for all presented calibrations and we need to discuss its influence on the photometric metallicity. The $m_2$ index  is a reddening independent parameter \citep[see e.g.][]{golay80}, thus we only have to consider the reference value of $m_2$ that is based on $(B2-V1)$. The flat part in Fig. \ref{fig:hyades} indicates that for early A to early F-type stars the derived metallicity is almost unaffected by the reddening. However, at $(B2-V1)=0.4$\,mag, the very cool end of the calibration, an uncertainty of 0.01\,mag in the colour owing to reddening (or the photometric measurement) corresponds to a metallicity shift of 0.06\,dex at solar metallicity, whereas at \FeH\ = $-$1\,dex we already have to consider an error that is twice as large. In the calibration procedure of $\Delta U_t$ both indices are affected by reddening, but the influence is somewhat compensated by the reddening vector and the larger baselines. For early G-type stars ($t \sim 0.20$\,mag) an error of 0.01\,mag in the Johnson \ubv\ colour excess $E(B-V)$ transforms into a shift of 0.05\,dex in the metal-poor range, and half the value in the metal-rich part. Down to early K-type stars the influence of reddening increases by a factor of two, whereas mid to late K-type objects are almost unaffected by reddening. We estimate that in the metal-poor range a change of the colour excess by 0.10\,mag results in a shift of only 0.05\,dex. We note that we adopt the reddening ratios listed by \citet{cramer99}.

The calibration samples might be influenced by reddening as well, in particular the stars farther than 50\,pc. We therefore checked these objects with the Str\"omgren-Crawford $uvby\beta$ system that is able to derive accurate reddening values \citep[see e.g.][]{karatas10}. We compiled photometric data from the catalogue by \citet{paunzen15}, only for 21 stars $uvby\beta$ measurements are not available. The intrinsic colour calibrations by \citet{napiwotzki93} and \citet{karatas10} were used to estimate $E(b-y)$ and we noticed only about 20 stars in each metallicity calibration sample that exceed $E(b-y) > 0.01$\,mag, but most are less reddened than 0.03\,mag. About half of the reddened objects in the $\Delta m_2$ sample are A- to early F-type stars, thus the influence of reddening is negligible. However, we tested the metallicity calibrations by correcting for the reddening and by excluding the few reddened objects, but noticed no significant effect on the results.

\begin{figure}
\centering
\resizebox{\hsize}{!}{\includegraphics{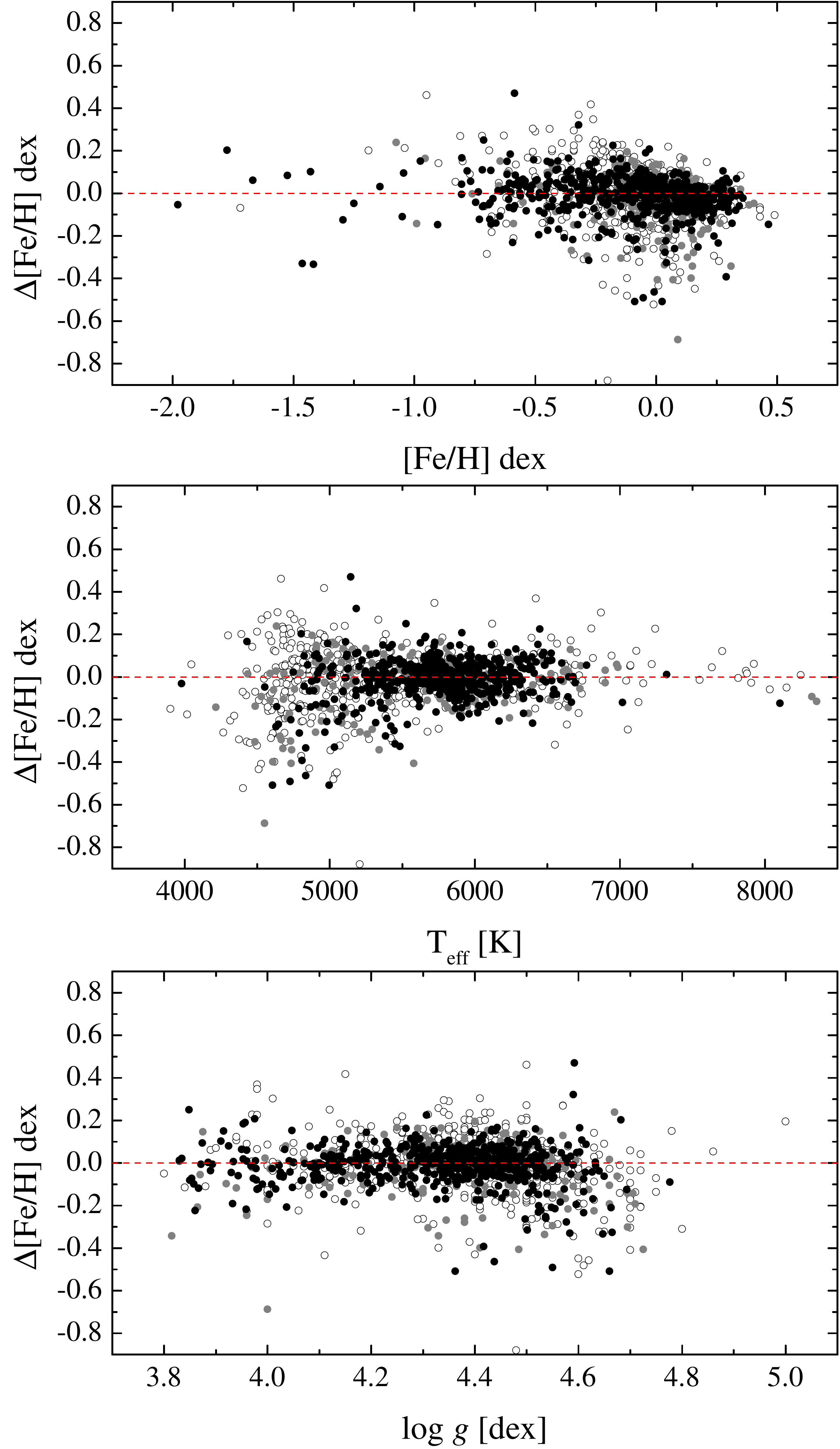}}
\caption{Differences of the photometric metallicities to the spectroscopic values ($\Delta$\FeH = phot $-$ spec) for the dwarfs closer than 50\,pc as a function of spectroscopic metallicity, effective temperature, and \logg. Black symbols are stars with  mean spectroscopic metallicities based on at least three determinations (thus in general the calibration sample), grey symbols objects with two spectroscopic measurements, and open symbols with single estimates.}
\label{fig:deltas}
\end{figure}

We apply the metallicity calibrations to the complete main sequence sample with spectroscopic metallicity determinations, and we adopt the $\Delta U_t$ calibration if the colour of an object allows the use of both calibrations. Figure \ref{fig:deltas} shows the differences to the spectroscopic values ($\Delta$\FeH) for stars closer than 50\,pc (to not overcrowd the figure) as a function of the spectroscopic abundance, effective temperature, and \logg. We derive linear regressions using the residuals of the calibration star sample and obtain the slopes $-$0.02 $\pm$ 0.01, +0.09 $\pm$ 0.08, and $-$0.01 $\pm$ 0.02 for the parameters \FeH, log(\Teff), and \logg. Note that also temperature was studied in the logarithmic space to avoid an expression in power of tens.  All slopes give no hint of a dependency which exceeds 2$\sigma$ or 0.05\,dex over the complete parameter range.    The complete samples (out to 150\,pc) with mean iron abundances (based on N = 2 and N $\geq 3$ measurements; 342 and 719 stars, respectively) show an accuracy of $\sigma \sim 0.11$\,dex, whereas the single measurement sample (898 objects) has a somewhat larger scatter ($\sigma \sim 0.14$\,dex). 

The middle panel of Fig. \ref{fig:deltas} shows that the scatter is much larger at lower temperatures. However, also the standard deviation of the mean spectroscopic abundances increases at lower temperatures, but also towards underabundant objects. Thus, if we restrict the N $\geq 3$ sample to stars hotter than 5200\,K and \FeH\ $> -1.0$\,dex, we obtain a better accuracy of $\sigma \sim 0.08$\,dex (602 objects).

\begin{figure}
\centering
\resizebox{\hsize}{!}{\includegraphics{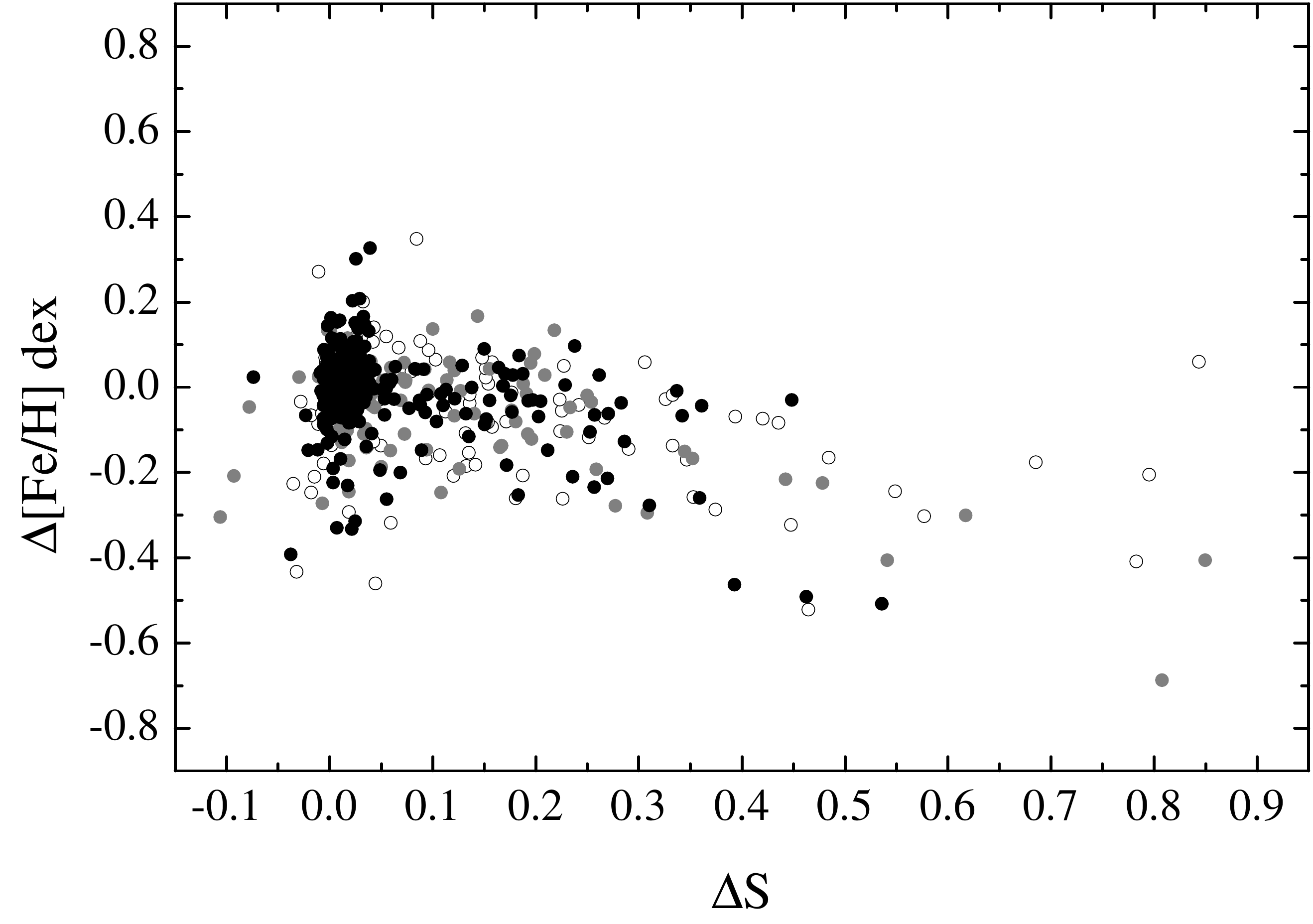}}
\caption{$\Delta \FeH$ as a function of the chromospheric activity level $\Delta$S, derived by \citet{isaacson10}. The symbols are the same as in Fig. \ref{fig:deltas}.}
\label{fig:chromo}
\end{figure}

We also observe a slight excess of cooler type stars for which the photometric metallicity is underestimated. This is certainly not caused by an unconsidered interstellar extinction, because most objects are reasonably close-by ($\lesssim$ 30\,pc). However, we notice that chromospheric activity influences the photometric metallicities. \citet{isaacson10} measured the flux in the core of the \ion{Ca}{ii} H and K lines of more than 2600 stars and define $\Delta S$ as an indicator of excess in emission. Figure \ref{fig:chromo} shows $\Delta \FeH$ as a function of $\Delta S$ for more than 500 objects in common. The stronger the emission the larger the influence on $\Delta U_t$ and the resulting photometric metallicity. Thus, in particular for very active G and K-type dwarfs, the metallicities should be used with caution.

Finally, we summarise the valid calibration ranges for dwarf stars. The $\Delta m_2$ calibration can be applied to stars with $-0.15 {\rm \,mag} \leq \Delta m_2  \leq 0.04 {\rm \,mag}$ or  $-2.0{\rm \,dex} \lesssim {\rm [Fe/H]} \lesssim 0.3 {\rm \,dex}$. The quadratic term of the calibration indicates that there is a saturation effect at higher metallicities. Thus, if the calibration is applied to metal-rich stars outside the valid range, we will certainly only obtain a lower limit of the metallicity. The $\Delta U_t$ calibration for cool type stars is a piecewise linear function with limits listed in Table \ref{tab:calib}. It can be applied to stars with metallicities up to [Fe/H] $\sim$ 0.5\,dex, but the data coverage suggests that [Fe/H] $\sim$ 0.4\,dex (or $\Delta U_t = 0.14$\,mag) is a more reasonable and secured upper limit.    

\subsection{Giant stars}
\label{giantcal}
The giant star sample was compiled by using the spectral types, the gravity ($\logg \lesssim 3.5$), and the absolute magnitude ($M_V < 2.0$\,mag). About 700 objects satisfy these criteria. For giant stars  we cannot rely on the Hyades (or another cluster) to define a reference sequence, because of the number of stars and the colour range that is covered. Thus, we have to use stars of the compiled sample itself for that purpose. As expected, the metallicity distribution shows a maximum around solar abundance. We therefore select 131 stars in the metallicity interval $-0.05 \leq \FeH \leq +0.05$ (with a median $\FeH = 0.00 \pm 0.03$\,dex) to define the reference sequence in the $t/U$ plane. In general, objects with at least two metallicity estimates were adopted, but towards the cool and hot temperature end we also selected lower quality results to obtain a better coverage. The reference sequence in the colour range $0.28 \leq t \leq 0.90$\,mag for early G to late K-type giants is shown in Fig. \ref{fig:hyades_gu} and listed in Table \ref{tab:sequence}.

There is a colour dependency on the blanketing parameter $\Delta U_t$ as well, and we correct that similarly as for the dwarfs, but here late G-type giants serve as the reference with almost identical corrections towards cooler and hotter stars. 
The final corrected sequence shows a linear dependency of $t$ and $U$, 
and a slope that reasonably agrees with the ones for G-type dwarfs. However, the two sequences significantly  disagree at the cool end, which is obviously a result of the increasing difference of gravity between dwarfs and giants of the same colour.

\begin{figure}
\centering
\resizebox{\hsize}{!}{\includegraphics{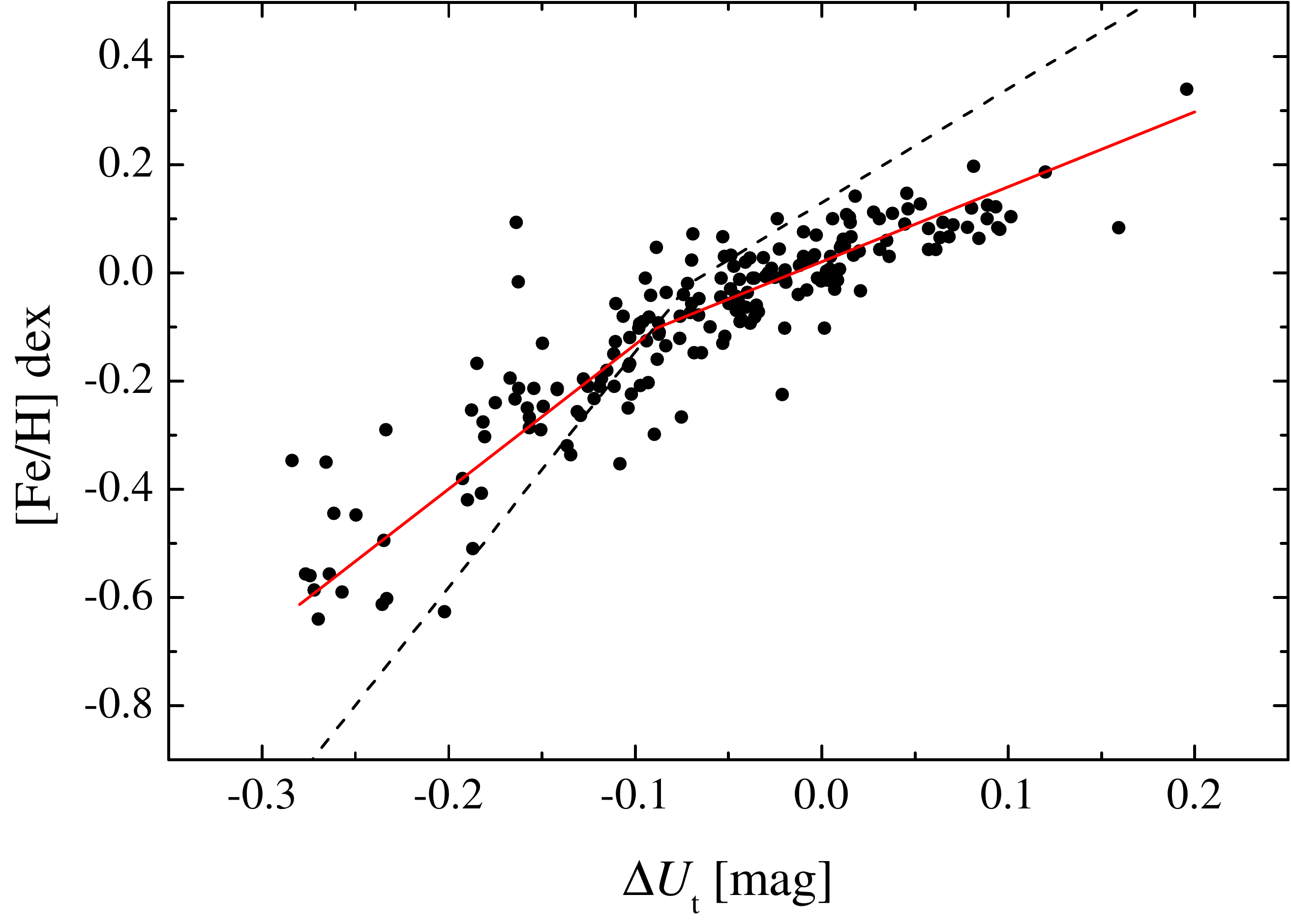}}
\caption{Calibration sample of 194 giant stars in the $\Delta U_t$/[Fe/H] plane. The red solid line shows the fit of a piecewise linear function as listed in Table \ref{tab:calib_gaint} and the dashed black line the fit to the dwarfs.}
\label{fig:calib_giant}
\end{figure}

For the metallicity calibration we adopt 205 stars that have not less than three iron abundance measurements, but for the final fit we exclude 11 objects with \logg\ $<$ 2.0\,dex as suggested by \citet{heiter14} and six objects that deviate more than 3$\sigma$. The sample and the piecewise linear fit are shown in Fig. \ref{fig:calib_giant}, and the coefficients of the fit are listed in Table \ref{tab:calib_gaint}. There is a lack of underabundant giants ($\FeH < -0.7$), which is in line with the findings by \citet{soubiran03} for the solar neighbourhood. However, the covered metallicity range is still well suited for a study of open clusters, for example.

\begin{table}
\caption{Metallicity calibration of giant stars.  } 
\label{tab:calib_gaint} 
\begin{center}
\begin{tabular}{l c c } 
\hline
 &  \FeH$_{\Delta U_t}$ &  \FeH$_{\Delta U_t}$ \\ 
 &   $-0.28 < \Delta U_t < -0.09$ & $-0.09 \leq \Delta U_t < 0.20$ \\
\hline 
$a_0$ &  0.14 (0.07)  & 0.02 (0.02) \\
$a_1$ &  2.69 (0.27)  & 1.39 (0.13)\\
N          &  63             &   125         \\
$\sigma$[dex]      & 0.08    &  0.06      \\

\hline 
\end{tabular}
\end{center}

\medskip
Notes: The regressions are given in the form a$_0$ + a$_1$ $\Delta U_t$. The errors are given in parentheses. The number of calibration stars in the respective colour range and the standard deviation of the metallicity calibrations are listed in the last lines.
\end{table}

The zero point of the calibration ($\FeH = 0.02$) is slightly above the mean metallicity of the standard sequence sample, but well within its scatter. Nonetheless, we do not fix the zero point to the solar value because the corrected sequence displaces the majority of stars towards lower $\Delta U_t$ values, resulting in a positive shift of the zero point. We note that the calibration of the dwarfs is unaffected  because the majority of stars are in the colour interval that requires almost no correction. 

The linear functions for the calibration of $\Delta U_t$ are somewhat flatter compared to the ones for dwarfs (cf. Fig. \ref{fig:calib_giant}). The linearity of the reference sequence already indicates that the influence of reddening is the same over the covered temperature range, and we derive that an error of 0.01\,mag in $E(B-V)$ corresponds to a change of \FeH\ by 0.03\,dex and 0.06\,dex for the metal-rich and metal-poor range, respectively.

\begin{figure}
\centering
\resizebox{\hsize}{!}{\includegraphics{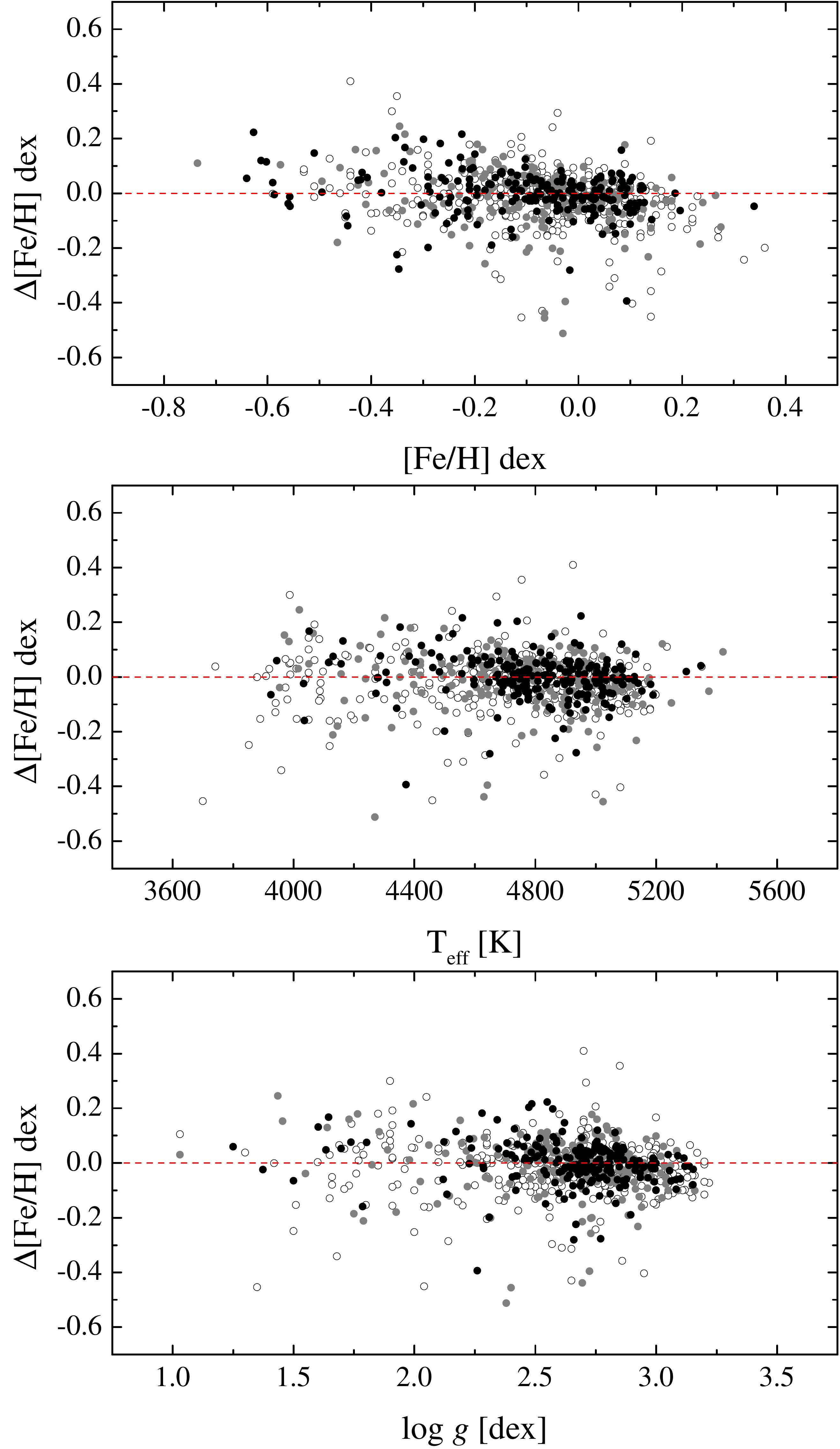}}
\caption{Differences of the photometric metallicities to the spectroscopic values ($\Delta$\FeH = phot $-$ spec) for all compiled giants as a function of spectroscopic metallicity, effective temperature, and \logg. The symbols are the same as in Fig. \ref{fig:deltas}.}
\label{fig:deltas_giants}
\end{figure}

Figure  \ref{fig:deltas_giants} shows $\Delta \FeH$ for all giants as a function of metallicity, temperature, and gravity. We derive a scatter of 0.08, 0.11, and 0.10\,dex for the subsamples in descending quality order. The upper panel of the figure might indicate that overabundant stars get systematically underestimated by the photometric metallicity, but most spectroscopic data are from the single measurement sample. We postpone an additional verification of the photometric scale to the analysis of open clusters in Sect. \ref{sect:clusters}. Otherwise, we do not notice a dependency that exceeds the typical scatter listed above. We apply linear regressions to the residuals of the calibration star sample and derive the slopes $-$0.05 $\pm$ 0.03, $-$0.24 $\pm$ 0.25, and $-$0.02 $\pm$ 0.02 for the parameters \FeH, log(\Teff), and \logg. Similar to the dwarf stars, there are no variations above 0.05\,dex over the complete parameter range and the significance of the slopes is below 2$\sigma$. However, the temperature dependency shows a much steeper slope and a larger error than the other parameters, but which is obviously caused by the narrow range in the logarithmic space.  We note that the calibration works well for stars with gravities lower than 2.0\,dex, which were excluded before, but they show a somewhat larger scatter ($\sigma = 0.10$\,dex). Though, the data indicate a marginal underestimation by $0.03 \pm 0.04$\,dex for the least evolved giants (\logg\ $\gtrsim$ 2.9). It is evident that this small correction is difficult to apply if spectroscopic data are not available. However, we tested the calibration also with a sample of subgiants that was in general selected based on the position in the colour-magnitude (CMD) and spectroscopic Hertzsprung-Russell diagram \citep{langer14}, because of the problems related to the accurate spectral classification of the luminosity class IV. We compiled about 80 objects, but most are from the lower quality spectroscopic sample. The calibration for the giant stars can be applied, but $\Delta U_t$ needs a small correction by $+0.025$\,mag to obtain well scaled results with $\sigma = 0.09$\,dex. This also corrects the small metallicity offset of the giants mentioned above. There are only a few A- to F-type giants and subgiants (27 objects), but the $\Delta m_2$ calibration for dwarfs works apparently well also for the early type giant stars without any additional correction, resulting in $\sigma = 0.10$\,dex. 

Finally, we also summarise the valid calibration range for the giant stars (cf. Table \ref{tab:calib_gaint}). They cover a more narrow metallicity range than the dwarfs as shown in Figure  \ref{fig:calib_giant}. The calibration is limited to [Fe/H] $\sim$ $-$0.6\,dex, and at present we cannot prove that an extrapolation to more underabundant objects is justified. There are only a few stars that anchor the metal-rich end of the calibration ($\Delta U_t \sim 0.2$\,mag or [Fe/H] $\sim$ 0.3\,dex), but in the next Section we verify the scale and show that an extrapolation up to [Fe/H] $\sim$ 0.4\,dex still provides reliable results.

\section{Additional comparisons of the metallicity scale}
\label{addcomp}
With our approach to homogenise the spectroscopic data in Sect. \ref{specmetal} we might have introduced an offset, thus additional comparisons are important. We therefore use the work by   \citet{jofre14}, which provides very accurate astrophysical parameters  for a sample of stars that should serve as calibrators for the \textit{Gaia} satellite mission, for example. They analysed 34 of such benchmark objects using several different methods. Certainly, too few stars for a direct calibration of the $Geneva$ photometry, but the sample covers a broad parameter space and is thus very suitable for an additional check of our results. 

\begin{figure}
\centering
\resizebox{\hsize}{!}{\includegraphics{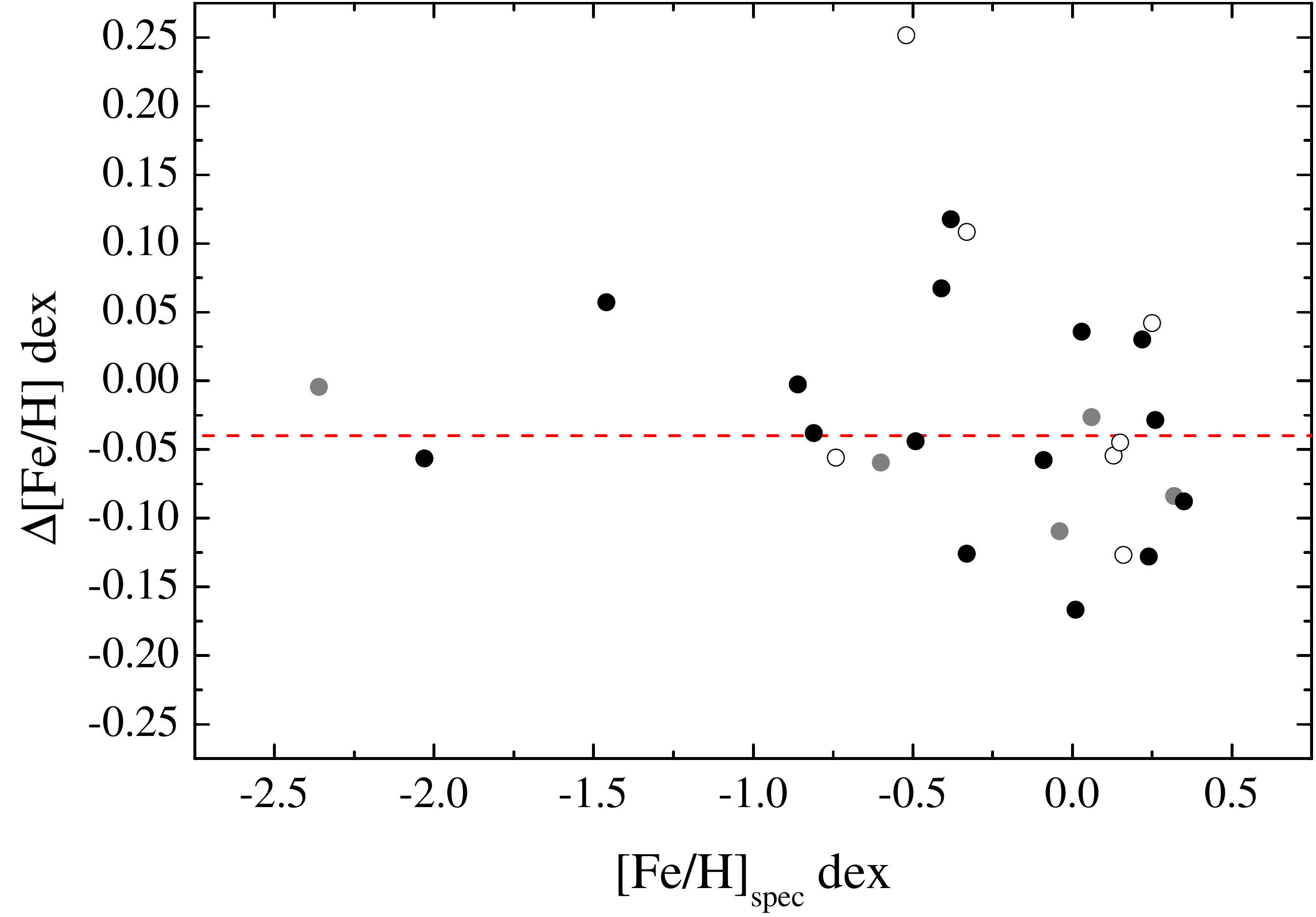}}
\caption{Comparison of the photometric metallicities with the spectroscopic results for the Gaia benchmark stars by \citet{jofre14}. The black symbols show dwarf stars, grey symbols the subgiants, and open circles giant stars. The dashed line shows the derived offset of $-$0.04\,dex.}
\label{fig:benchmark}
\end{figure}

We subdivide the stars into dwarfs, subgiants, and giants as listed by \citet{heiter15} and applied the corresponding calibration presented here. We derived photometric metallicities for almost all stars from the list other than the five M-type giants, the metal-poor giant HD~122563 (it is outside the calibration range), and the Sun. Figure \ref{fig:benchmark} shows the comparison with the spectroscopic benchmark metallicities. The largest deviation is found for HD~124897 (Arcturus) with an overestimation by 0.25\,dex. Arcturus is one of the very brightest stars in the $Geneva$ catalogue, thus problems with the photometric standardisation are possible. Nonetheless, we obtain $\sigma = 0.09$\,dex using all stars, and the exclusion of Arcturus results in a median difference between the two scales of $\Delta \FeH = -0.04 \pm 0.07$\,dex. The difference does not significantly differ for the individual subgroups and photometric calibrations. However, the offset might be explained by the spectroscopic value for the Sun, \citet{jofre14} list an overabundance by almost the same amount ($\FeH = +0.03 \pm 0.01$\,dex). 

\begin{figure}
\centering
\resizebox{\hsize}{!}{\includegraphics{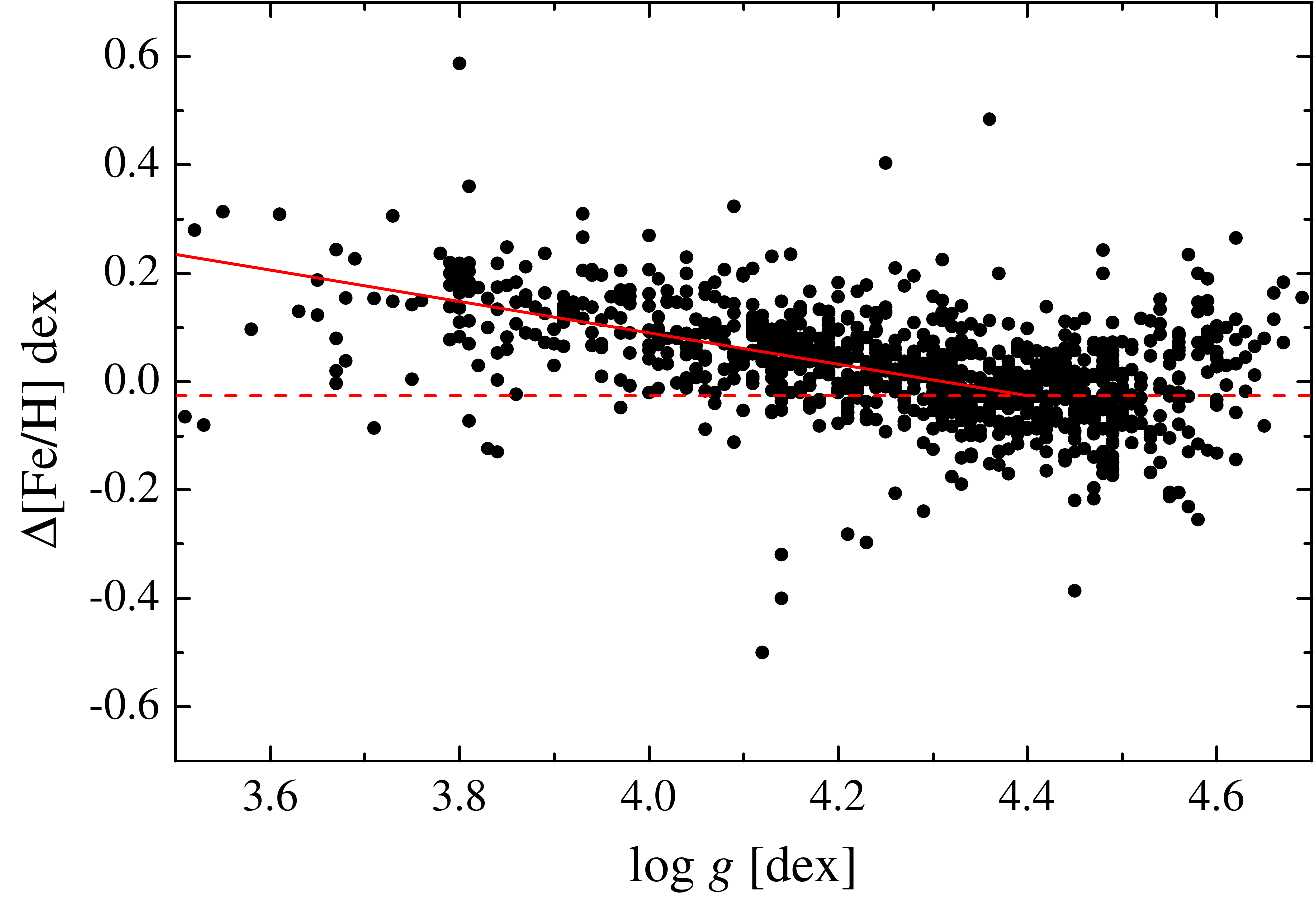}}
\caption{Comparison of the GCS metallicities with mean spectroscopic data as a function of gravity. The solid line shows the linear relationship for stars with \logg\ $<$ 4.4\,dex as presented in Equation \ref{loggdep}  and the dashed line shows the small offset for stars with higher gravities. }
\label{fig:gravdep}
\end{figure}

Finally, we compare the $Geneva$ metallicities with another set of photometric measurements. \citet{casagrande11} re-analysed the data of the Geneva-Copenhagen survey (GCS), which is based on Str\"omgren photometry for about 17\,000 F- and G-type dwarfs in the solar neighbourhood. Thus, we can expect a very large overlap with our samples. We compare their results with about 1000 objects from our sample of spectroscopic calibrators (N $\geq$ 3 measurements) and confirm the conclusions by \citet{casagrande11} that there are no significant dependencies on effective temperature or metallicity. However, they have not presented a comparison of the metallicity scale as a function of \logg , but Fig. \ref{fig:gravdep} clearly shows that the GCS metallicities of more evolved objects are overestimated by up to about 0.2\,dex, and we derive the following linear relation to correct the results for stars with \logg\ $<$ 4.4\,dex, whereas for stars with higher gravities we find that the GCS metallicities are underestimated by the small amount of $-0.03 \pm 0.09$\,dex. 
\begin{equation}
\label{loggdep}
\FeH  = \FeH_{\rm GCS} - \left(1.25^{\pm0.06}-0.29^{\pm0.01}\logg\right) 
\end{equation}

We note that we adopt the GCS gravities, which agree well with other published results. We obtain $\Delta \FeH = +0.00 \pm 0.08$\,dex (GCS $-$ spec) after the correction and the exclusion of a few 3$\sigma$ outliers. The accuracy of the corrected data has improved compared with $\sigma = 0.097$\,dex, the result by \citet{casagrande11}. Furthermore, a direct comparison of the two photometric scales shows that these are equally scaled with $\sigma = 0.11$\,dex.

We derive a mean photometric metallicity based on our data and GCS and compare these again with the calibrators. The scatter improves to $\sigma = 0.06$\,dex, thus the combination of the two photometric systems apparently reduces the individual limitations, and we will make use of this benefit in Sect. \ref{sect:photsamples}. 

\subsection{Open cluster metallicities}
\label{sect:clusters}
Open cluster metallicities are important to improve our knowledge of the Milky Way's chemical evolution, but can also serve as an additional benchmark of the new calibrations. Recently, \citet{netopil16} presented homogenised metallicities of 172 open clusters in total based on spectroscopic and photometric estimates, and investigated radial migration effects, for example. 
From this work and the additional literature, it is evident that the $Geneva$ photometric system was not yet systematically used to derive open cluster metallicities. We therefore queried the open cluster database WEBDA\footnote{http://webda.physics.muni.cz} \citep{mermilliod03} and the GCPD for objects that are covered by $Geneva$ data. 

The CMDs in the $Geneva$ system, for additional guidance also in other colours such as in \ubv, were checked to divide the stars into dwarfs, subgiants, and giants. 
The magnitude limit of the photometric data ($\sim$\,12\,mag) restricts the open cluster sample to close-by objects, thus available proper motions (e.g. Tycho-2, $Gaia$) already provide an accurate membership criterion. Parallax measurements in the first data release of the $Gaia$ satellite mission \citep{gaia2016} provide an additional valuable membership information. Furthermore, we used membership probabilities listed in WEBDA and the additional literature, combined with an evaluation of the photometric colours. In particular the low number of giants makes a proper member selection important, though a membership analysis based on radial velocities \citep{mermilliod08} is available for almost all compiled giants as well.  

We derive metallicities for 54 open clusters in total (30 based on dwarfs and 32 based on giant stars) and therefore it is one of the largest samples based on a homogeneous photometric calibration \citep[see discussion by][]{netopil16}. There are eight objects for which we derive metallicities based on dwarfs and giants. These agree very well within the errors and deviate by not more than 0.02\,dex, providing an additional proof that the two metallicity scales are consistent.

The photometric metallicity of the Hyades based on dwarf stars cannot be considered as an independent measure because the standard sequences and the zero point rely on this cluster. However, we derive \FeH\ = +0.13 $\pm$ 0.04\,dex, a scatter which is only twice as large as recently found by \citet{liu16} in a detailed abundance analysis of 16 solar-type stars in the cluster. They attribute the scatter to an intrinsic
abundance dispersion. For the four giant stars in the cluster we derive \FeH\ = +0.09 $\pm$ 0.05\,dex, though \citet{dasilva15} list \logg\ $\gtrsim$ 3.0 for all stars. Thus, with the correction discussed in Sect. \ref{giantcal} the mean metallicity is very close  to the result for the dwarfs.

Both dwarf star calibrations ($\Delta m_2$ and $\Delta U_t$) are applicable to two additional nearby clusters: NGC~2632 (Praesepe) and Melotte~22 (the Pleiades).
For Praesepe we derive \FeH\ = +0.15 $\pm$ 0.07\,dex (65 stars) and +0.14 $\pm$ 0.04\,dex (25 stars) using $\Delta m_2$ and $\Delta U_t$, respectively. The Pleiades show a metallicity very close to solar: \FeH\ = +0.00 $\pm$ 0.08\,dex (49 stars) and $-$0.02 $\pm$ 0.07\,dex (11 stars) based on $\Delta m_2$ and $\Delta U_t$. The calibrations result in good agreement, thus for these two clusters Table \ref{tab:metaldata} lists the mean values using all data for dwarf stars.

A proper choice of the cluster reddening is important to derive reliable metallicity values. For 29 open clusters, we estimate the reddening directly with the $Geneva$ data and the reddening free $X/Y$ indices of B-type cluster stars \citep[see e.g.][]{cramer99}. To avoid a confusion owing to different colour excesses, we always list the more common $E(B-V)$ values, which were obtained with the reddening ratio given by \citet{cramer99}. The reddening slightly differs as a function of intrinsic colour. Thus, if the reddening was derived for early type main sequence stars, a small correction for giant stars is necessary. We therefore downscale such results by 10\,\% as discussed by \citet{twarog97}. The last reference evaluated the reddening values of open clusters for main sequence stars and red giants, and we adopt their values for 15 clusters. For one cluster (IC~4756) they note a variable reddening for the giant stars and do not list a value. We therefore estimate it using Str\"omgren-Crawford data of subgiants, but also confirm their result for the main sequence. We also use some results by \citet{netopil13} that are based on the Str\"omgren-Crawford photometric system and the $Q$-method in the \ubv\ system. However, for a few little studied open clusters, we have to refer to other references or estimate the reddening using available spectroscopic temperatures of giant stars and the empirical temperature calibration by \citet{huang15}. However, a small error or offset of 50\,K ($\sim$\,1\%) in the effective temperature already results in a deviating reddening scale by about 0.03\,mag, which has to be added to the uncertainties and the scale of the calibration itself and the photometric errors. We note that different line-lists might cause temperature differences of more than 100\,K \citep[see e.g.][]{santos09}. A check of their results for giants in open clusters and a comparison of the derived reddening with photometric determinations shows that for the giant stars optimised line-list by \citet{hekker07} provides better results.  

Table \ref{tab:metaldata} lists the respective choice of the reddening, and if better reddening estimates become available for a cluster, a rough correction of the metallicities is possible with the values of the reddening influence that are discussed in Sects. \ref{dwarf_cal} and \ref{giantcal}. However, most cluster metallicities based on dwarfs and the $\Delta m_2$ calibration were derived using A- and early F-type stars, thus they are little affected by the interstellar reddening. Nonetheless, we notice that these metallicities show in general a larger scatter than the results based on giant stars. The larger baseline of $\Delta U_t$ probably  lowers the influence of variability, binarity, or rotation, to mention a few properties that might affect the photometric colours. 

\begin{figure}
\centering
\resizebox{\hsize}{!}{\includegraphics{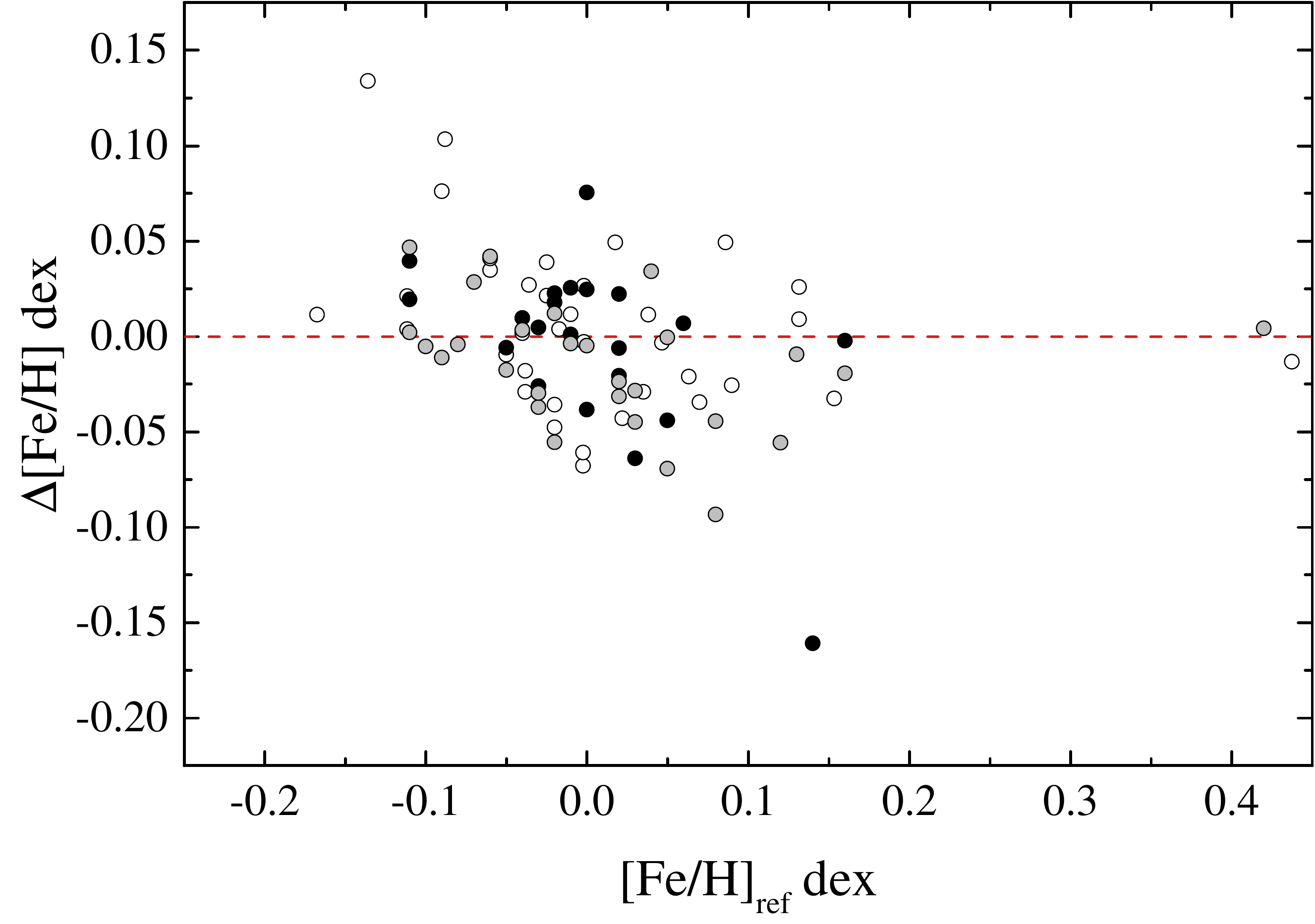}}
\caption{Comparison of the $Geneva$ open cluster metallicities with the homogenised reference scale by \citet{netopil16} and the additional literature. The black and grey circles show the differences to the high-quality spectroscopic data, divided into the $Geneva$ results for dwarf stars and subgiants/giants, respectively. Open circles show the comparison with mean photometric metallicities. }
\label{fig:clustercomp}
\end{figure}

As already mentioned, the open cluster metallicities can be used as an additional check of the calibrations. The comparison of data for 42 objects (Fig. \ref{fig:clustercomp}) that have high-quality spectroscopic results \citep[in respect of $S/N$ and spectral resolution, see][]{heiter14,netopil16} shows  very good agreement. The comparison also includes some more recent spectroscopic results for NGC~2547, NGC~2451A, and NGC~2451B by \citet{spina17} based on UVES data, for NGC~5316 by \citet{drazdauskas16}, and for NGC~6940 by \citet{bocek16}. The largest deviation in the comparison was found for Melotte~20 (Alpha Persei), the $Geneva$ metallicity $\FeH\ = -0.02 \pm 0.11$ is underestimated by 0.16\,dex. However, the adopted spectroscopic metallicity value for the cluster shows a large error of 0.11\,dex and is only based on two stars. This might be caused by too strict selection criteria introduced by \citet{heiter14}. The mean values of the two available studies for the cluster without any restriction on the data \citep[$-$0.02 and $-$0.05\,dex, Table 4 by][]{heiter14} are in good agreement with the estimate presented here and other photometric determinations. The exclusion of this data point results in a median $\Delta \FeH\ = +0.01 \pm 0.03$\,dex (20 clusters) based on dwarf star metallicities and $-0.01 \pm 0.03$\,dex (28 objects) for the estimates using giant stars. The comparison with mean photometric values listed by \citet{netopil16} includes only  estimates based on at least two photometric systems (36 objects) and yields $\Delta \FeH\ = +0.00 \pm 0.04$\,dex ($Geneva -$ others). The largest differences in this comparison are found for IC~2391 and NGC~2547. For both clusters the Str\"omgren data by \citet{lynga84} apparently distort the mean values listed by \citet{netopil16} to lower metallicities. However, for the two objects spectroscopic results are already available, which agree very well with the $Geneva$ data.

We conclude that the $Geneva$ metallicities are very well scaled, and the low scatter also provides a hint for the in general good accuracy of the adopted reddening values. Furthermore, the result for NGC~6791, one of the most metal-rich open cluster, shows that the metallicity calibration of the giants can be applied up to at least +0.40\,dex. The agreement with the spectroscopic metallicity suggests that the strong underabundant CN to Fe ratios in this cluster do not influence the result as in other photometric systems \citep[e.g. DDO, see discussion by][]{netopil16}.

For one open cluster (NGC~6067) we do not provide a metallicity estimate, although ten giant star members \citep{mermilliod08} are covered by $Geneva$ data. A very large scatter in $\Delta U_t$ was found, and most values are outside the calibration range. This cannot be explained even by assuming strong differential reddening. The object is located in a very crowded field, thus the photoelectric data are most probably affected by additional stars in the aperture.   

Table \ref{tab:metaldata} suggests that there is no other metallicity value available for two clusters (NGC~2281 and NGC~6025). However, the compilation of photometric metallicities by \citet{paunzen10} lists estimates based on the \ubv\ colours, which were excluded by \citet{netopil16} for reasons discussed therein. The $Geneva$ and \ubv\ data agree for NGC~2281, but differ by about 0.3\,dex for the other object. 
Though, \citet{cameron85} used photographic measurements for the analysis of this object, which are probably too inaccurate. Following the procedure by \citet{netopil16}, we finally present an update of weighted mean photometric metallicities based on their list and the new determinations. For the eight clusters with metallicities based on dwarf stars and giants a mean value was adopted. For most objects of the sample spectroscopic data are already available, but the new photometric data provide an independent measure for them and improved the reliability of photometric estimates for the remaining not yet spectroscopically studied open clusters.

\section{Catalogue of photometric metallicities}
\label{sect:photsamples}

In Sect. \ref{addcomp} we noticed that a combination of the $Geneva$ metallicities with the ones of the GCS results in a reasonable improvement of accuracy ($\sigma$ $\sim$ 0.06\,dex). Thus, such a merged sample can be useful in many respects, the data can serve as primary or secondary calibrators of other data sets, for example. The advantage is certainly that both results are on a homogeneous scale and cover many objects.

We therefore performed a match of the $Geneva$ catalogue with the GCS, resulting in an initial list of almost 5000 stars. Though, for numerous objects the GCS calibration was not applied within its range, and we exclude these stars using their flag as suggested by \citet{casagrande11}. Stars without a \logg\ information in the GCS were removed from the list as well, because of the dependency of these metallicities on the gravity as discussed in Sect. \ref{addcomp}; we correct the remaining data accordingly. Furthermore, we use the effective temperatures of the GCS to exclude stars with \Teff\ $<$ 5200\,K because of the reasons discussed in Sect. \ref{dwarf_cal}. We note that the GCS metallicities also show a larger scatter for cooler stars and that chromospheric activity influences their results as well, though to a somewhat smaller extent (about 0.2\,dex for the apparently most active stars). However, for the sake of reliability, we suggest to adopt even a more conservative lower temperature limit of 5500\,K. Chemically peculiar stars are generally rare in the covered temperature range, we thus need to exclude only a few of them.   

The final catalogue includes 3267 dwarf stars, for which we derive the metallicities in the $Geneva$ system as outlined in Sect. \ref{dwarf_cal} and by adopting the GCS reddening values. We note that the sample includes several binaries with individually measured components or combined photometric measurements. For these stars we used the remarks about the duplicity in both catalogues, but also the magnitudes in the two photometric systems to assure a proper match, taking into account that discrepancies in the designation of binary components are likely. The interested reader should keep this in mind if using these data. 

\begin{figure}
\centering
\resizebox{\hsize}{!}{\includegraphics{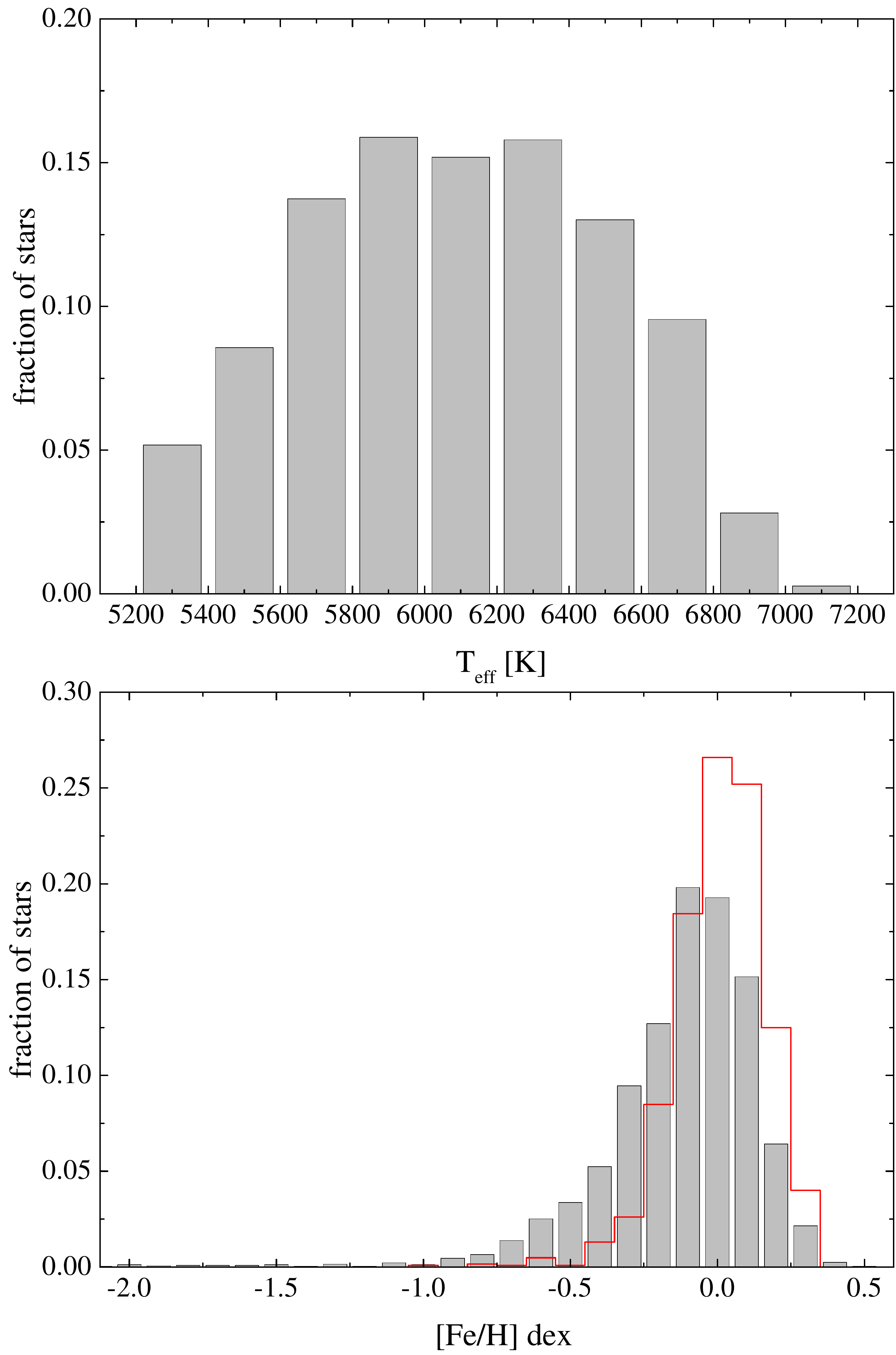}}
\caption{Distribution of the effective temperature and metallicity for the stars in the compiled sample of mean photometric metallicities (grey bars). The red line denotes the metallicity distribution of the A to F-type star sample. }
\label{fig:sampledistri}
\end{figure}

Figure \ref{fig:sampledistri} shows the distribution of the sample for the effective temperature and metallicity. The latter covers a broad range from about $-$2.0\,dex to overabundant objects with almost +0.5\,dex. The standard errors of the mean metallicities do not exhibit a dependency on temperature, gravity, or metallicity. Furthermore, more than two third of the stars show an error of the mean metallicity of $\leq$\,0.05\,dex, representing an additional proof that the two selected data sets agree on a very good level.  

We noticed another big advantage of the metallicity calibrations in the previous sections: the results for dwarf stars and giants in the spectral type range from about A3 to F2 are almost unaffected by the reddening. This temperature range ($\sim$\,6800\,K$-$8500\,K) is not or only partly covered by the sample defined above, but also spectroscopic results are rare. The sample of modern spectroscopic data, which we compiled in Sect. \ref{specmetal}, includes only about 650 objects in this temperature range and stem from numerous individual references. About 20 per cent of the stars belong to chemically peculiar groups, and a similar percentage is located in the Kepler field. Furthermore, there are concentrations of stars in open cluster areas (such as the Hyades or Praesepe) as well, resulting in a strong excess of measured northern stars ($\sim$\,70\%). Thus, there is only a very limited number of normal type stars on a homogeneous metallicity scale that are ``randomly'' distributed over the sky, usable to check the metallicity scale of new data for example. 
 
\begin{figure}
\centering
\resizebox{\hsize}{!}{\includegraphics{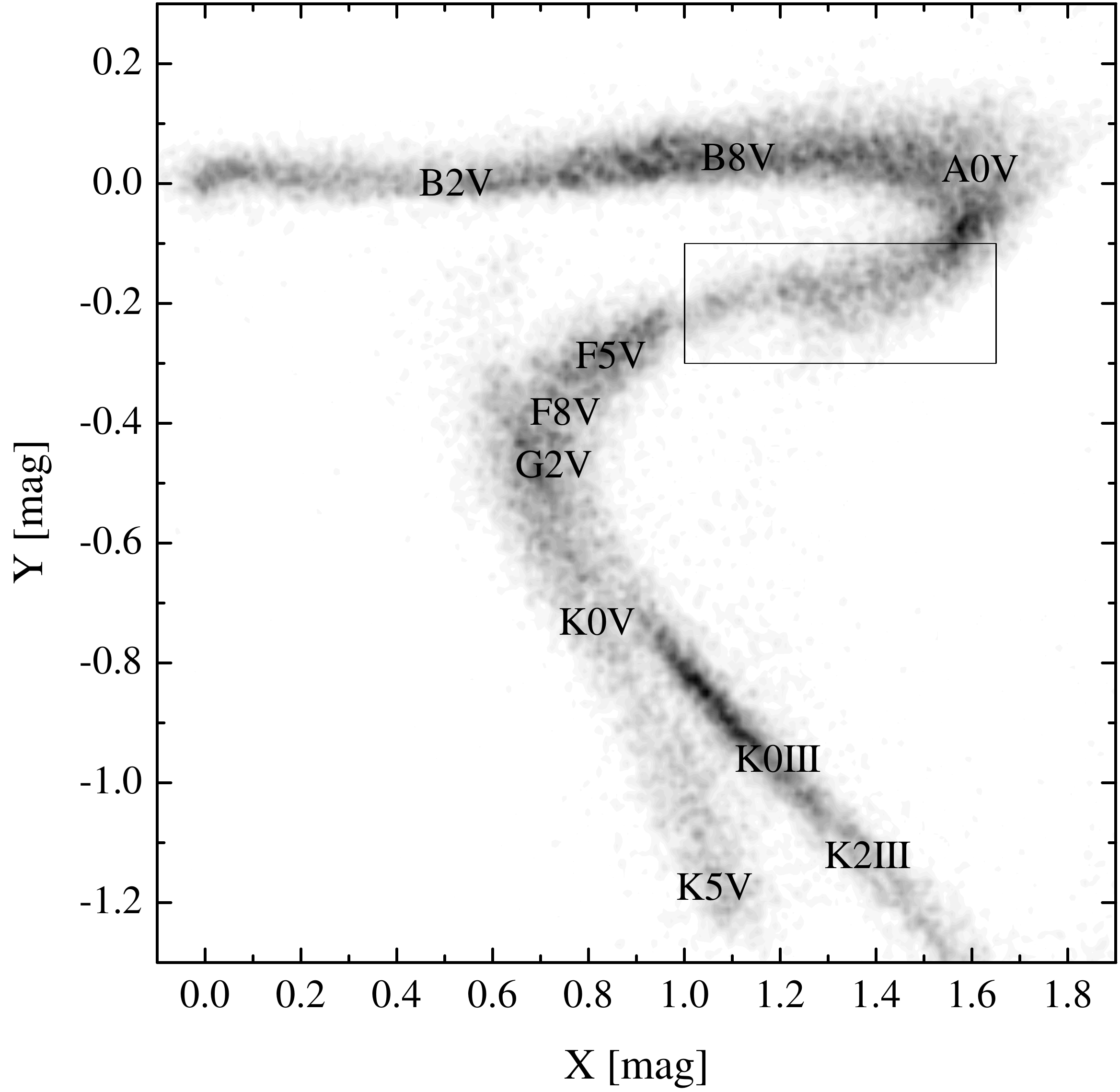}}
\caption{Distribution of the stars in the $Geneva$ catalogue in the reddening free X/Y two-colour diagram. Some spectral types are listed for guidance and the rectangle highlights the selected sample of A3 to F2 stars. }
\label{fig:densityplot}
\end{figure}

We therefore make use of the reddening free $X/Y$ indices in the $Geneva$ system and the intrinsic colours of MK spectral types by \citet{hauck94} to select an initial sample of early A to F-type field stars from the complete $Geneva$ catalogue using $1.0 \leq X \leq 1.65$ and $-0.3 \leq Y \leq -0.1$. Figure \ref{fig:densityplot} shows the stars in $Geneva$ catalogue in this two-colour diagram and the range of the adopted sample. We compiled spectral types for these stars to exclude objects that are supposed to be too hot or too cool for our purpose, and chemically peculiar stars were removed from the list as well. We also exclude the small overlap with our catalogue of mean metallicities, though the present list still includes stars from the GCS -- those which are flagged to be outside of the calibration range.  

The final list includes 1226 objects, for which we derive the metallicity by adopting $m_2$(std)=$-0.480\pm0.003$\,mag, the mean value for stars in the Hyades with the colours $-0.065 \leq (B2-V1) \leq 0.160$ (cf. Fig. \ref{fig:hyades}), and the procedure discussed in Sect. \ref{dwarf_cal}. However, the multivariate regression in Table \ref{tab:calib} also depends on the colour term $(B2-V1)_0$. The influence gets negligible towards metal-rich stars (cf. also Fig. \ref{fig:calibration}), but simply adopting a mean colour for the complete sample might introduce an offset of up to 0.06\,dex for the very coolest and hottest stars with underabundances around \FeH\ $\sim$ $-$0.5. For 238 GCS stars in the sample we can directly adopt the reddening estimates of the survey to correct the colours, but Str\"omgren data are available in the catalogue by \citet{paunzen15} for another 636 objects. These data were used to estimate the reddening as outlined in Sect. \ref{metalcal}. For the remaining stars we use the intrinsic colours by \citet{hauck94} and the compiled spectral types to estimate the intrinsic colours of the objects, and we estimate an accuracy of 0.03\,mag by a comparison to the other results. We consider this sample as a valuable extension of the first catalogue to hotter stars.

Figure \ref{fig:sampledistri} shows the metallicity distribution of the sample, which differs compared with the previous one of mean metallicities for cooler type stars. Obviously, there is a higher percentage of metal-rich objects. The low number of about 40 ``hot'' stars in the calibration sample might introduce an offset of the metallicity scale. However, the results for the open clusters do not suggest an improper calibration. The $Geneva$ metallicities of NGC~2547 and NGC~6405, for example, are based only on stars in the colour range that we adopt here and agree very well with spectroscopic data (cf. Table \ref{tab:metaldata}). We note that the result by \citet{kilicoglu16} for the latter object was not considered in the comparison in Sect. \ref{sect:clusters} owing to the adopted  criteria. Furthermore, the open cluster Melotte~20 covers several stars cooler and hotter than $(B2-V1)=0.16$\,mag and their mean values agree within 0.02\,dex. We therefore attribute the disagreement of the two metallicity distributions to the colour cuts and stellar evolution; see for example the discussion and figure 7 by \citet{bergemann14} or \citet{casagrande11} for a comparable effect. 

Both catalogues will be made available in electronic form at the CDS and provide all basic information (ID's and coordinates), colours that allow reproducing all the results, and finally the estimated (mean) metallicities.

\section{Conclusions}
\label{sect:conclusion}

We presented new metallicity calibrations for the $Geneva$ photometric system, both for dwarf stars and giants. The calibrations are applicable to a wide range of spectral types, from early A- to K-type stars, and the external accuracy is in most cases better than 0.10\,dex. The high homogeneity of the photometric system and a high number of covered stars makes it a valuable independent source to check the metallicity scale of different data sets.

We compared the derived metallicities with some  spectroscopic and photometric data and found in general very good agreement. However, compared to the $Gaia$ benchmark stars by \citet{jofre14} we identified a small offset of about 0.04\,dex that might be explained by their overabundant result for the Sun. The calibrations show a comparable accuracy as the Geneva-Copenhagen survey by \citet{casagrande11} that is based on Str\"omgren-Crawford photometry. Here we noticed an overlooked dependency of metallicity on gravity and we provide correction terms. However, a combination of $Geneva$ and Str\"omgren data results in some benefits: Str\"omgren photometry provides accurate reddening estimates also for cooler type stars, and the derived metallicities in these two systems can be compared and averaged to improve the accuracy. Therefore, we provide a list of mean photometric metallicities for about 3300 stars in the temperature range of 5200\,K to about 7000\,K that are expected to show a good external accuracy ($\sigma = 0.06$\,dex).

Another big advantage of the $Geneva$ metallicities is certainly the almost reddening free characteristic for early A to early F-type stars, a temperature range that is little covered by spectroscopic data \citep[see e.g.][]{soubiran16}. We therefore derived metallicities for more than 1200 stars to extend the previous list to hotter stars (up to $\sim$\,8500\,K).

Furthermore, we applied the metallicity calibrations to open cluster data. These provide on the one hand an additional proof of the metallicity scale and accuracy if compared to mean spectroscopic results, but also improves the knowledge of metallicity for open clusters that are not studied spectroscopically by now. In total, we present metallicities of 54 open clusters, 10 of them without any spectroscopic result. The  comparison with available spectroscopic metallicities shows very good agreement with a scatter as low as 0.03\,dex, but we also identified a some probably erroneous open cluster results that were adopted by a recent compilation.

We already identified about 20 spectroscopic data sets in the calibration star sample that show offsets up to about 0.4\,dex. The presented large catalogues of (mean) photometric metallicities certainly provide  large overlaps with other already available spectroscopic results or upcoming data, such as from the $Gaia$ satellite mission. Thus, the catalogues are suitable as primary or secondary calibrators to identify possible offsets and to merge all spectroscopic data into a homogenised database of stellar metallicities. 

The photoelectric $Geneva$ data, which are used for the metallicity calibrations, are certainly a closed box. However, attempts to secure $Geneva$ photometry also in the CCD era are presented for example by \citet{raboud97} with a study of the young cluster NGC~6231. More recently, \citet{mowlavi13} used a subset of the $Geneva$ filters for a variability study of the stars in NGC~3766. The authors note, however, that a more detailed investigation is still needed to closely tie the CCD $Geneva$ system to the standard one. This delicate standardisation along with observations in all $Geneva$ filters is of importance to make use of all the capabilities of the system, but also to apply the metallicity calibrations on new CCD data.

\section*{Acknowledgements}

MN acknowledges the support by the grant 14-26115P of the Czech Science Foundation GA\,\v{C}R. This research has made use of the WEBDA database, operated at the Department of Theoretical Physics and Astrophysics of the Masaryk University.




\bibliographystyle{mnras}
\bibliography{geneva_ref} 

\begin{thebibliography}{}
\makeatletter
\relax
\def\mn@urlcharsother{\let\do\@makeother \do\$\do\&\do\#\do\^\do\_\do\%\do\~}
\def\mn@doi{\begingroup\mn@urlcharsother \@ifnextchar [ {\mn@doi@}
  {\mn@doi@[]}}
\def\mn@doi@[#1]#2{\def\@tempa{#1}\ifx\@tempa\@empty \href
  {http://dx.doi.org/#2} {doi:#2}\else \href {http://dx.doi.org/#2} {#1}\fi
  \endgroup}
\def\mn@eprint#1#2{\mn@eprint@#1:#2::\@nil}
\def\mn@eprint@arXiv#1{\href {http://arxiv.org/abs/#1} {{\tt arXiv:#1}}}
\def\mn@eprint@dblp#1{\href {http://dblp.uni-trier.de/rec/bibtex/#1.xml}
  {dblp:#1}}
\def\mn@eprint@#1:#2:#3:#4\@nil{\def\@tempa {#1}\def\@tempb {#2}\def\@tempc
  {#3}\ifx \@tempc \@empty \let \@tempc \@tempb \let \@tempb \@tempa \fi \ifx
  \@tempb \@empty \def\@tempb {arXiv}\fi \@ifundefined
  {mn@eprint@\@tempb}{\@tempb:\@tempc}{\expandafter \expandafter \csname
  mn@eprint@\@tempb\endcsname \expandafter{\@tempc}}}

\bibitem[\protect\citeauthoryear{{Anthony-Twarog}, {Deliyannis}  \&
  {Twarog}}{{Anthony-Twarog} et~al.}{2016}]{anthony-twarog16}
{Anthony-Twarog} B.~J.,  {Deliyannis} C.~P.,   {Twarog} B.~A.,  2016, \mn@doi
  [\aj] {10.3847/0004-6256/152/6/192}, \href
  {http://adsabs.harvard.edu/abs/2016AJ....152..192A} {152, 192}

\bibitem[\protect\citeauthoryear{{Bensby}, {Feltzing}  \& {Oey}}{{Bensby}
  et~al.}{2014}]{bensby14}
{Bensby} T.,  {Feltzing} S.,   {Oey} M.~S.,  2014, \mn@doi [\aap]
  {10.1051/0004-6361/201322631}, \href
  {http://adsabs.harvard.edu/abs/2014A%26A...562A..71B} {562, A71}

\bibitem[\protect\citeauthoryear{{Bergemann} et~al.,}{{Bergemann}
  et~al.}{2014}]{bergemann14}
{Bergemann} M.,  et~al., 2014, \mn@doi [\aap] {10.1051/0004-6361/201423456},
  \href {http://adsabs.harvard.edu/abs/2014A%26A...565A..89B} {565, A89}

\bibitem[\protect\citeauthoryear{{Berthet}}{{Berthet}}{1990}]{berthet90}
{Berthet} S.,  1990, \aap, \href
  {http://adsabs.harvard.edu/abs/1990A%26A...236..440B} {236, 440}

\bibitem[\protect\citeauthoryear{{B{\"o}cek Topcu}, {Af{\c s}ar}  \&
  {Sneden}}{{B{\"o}cek Topcu} et~al.}{2016}]{bocek16}
{B{\"o}cek Topcu} G.,  {Af{\c s}ar} M.,   {Sneden} C.,  2016, \mn@doi [\mnras]
  {10.1093/mnras/stw1974}, \href
  {http://adsabs.harvard.edu/abs/2016MNRAS.463..580B} {463, 580}

\bibitem[\protect\citeauthoryear{{Bond}, {Tinney}, {Butler}, {Jones}, {Marcy},
  {Penny}  \& {Carter}}{{Bond} et~al.}{2006}]{bond06}
{Bond} J.~C.,  {Tinney} C.~G.,  {Butler} R.~P.,  {Jones} H.~R.~A.,  {Marcy}
  G.~W.,  {Penny} A.~J.,   {Carter} B.~D.,  2006, \mn@doi [\mnras]
  {10.1111/j.1365-2966.2006.10459.x}, \href
  {http://adsabs.harvard.edu/abs/2006MNRAS.370..163B} {370, 163}

\bibitem[\protect\citeauthoryear{{Brewer} \& {Carney}}{{Brewer} \&
  {Carney}}{2006}]{brewer06}
{Brewer} M.-M.,  {Carney} B.~W.,  2006, \mn@doi [\aj] {10.1086/498110}, \href
  {http://adsabs.harvard.edu/abs/2006AJ....131..431B} {131, 431}

\bibitem[\protect\citeauthoryear{{Cameron}}{{Cameron}}{1985}]{cameron85}
{Cameron} L.~M.,  1985, \aap, \href
  {http://adsabs.harvard.edu/abs/1985A%26A...147...39C} {147, 39}

\bibitem[\protect\citeauthoryear{{Casagrande}, {Sch{\"o}nrich}, {Asplund},
  {Cassisi}, {Ram{\'{\i}}rez}, {Mel{\'e}ndez}, {Bensby}  \&
  {Feltzing}}{{Casagrande} et~al.}{2011}]{casagrande11}
{Casagrande} L.,  {Sch{\"o}nrich} R.,  {Asplund} M.,  {Cassisi} S.,
  {Ram{\'{\i}}rez} I.,  {Mel{\'e}ndez} J.,  {Bensby} T.,   {Feltzing} S.,
  2011, \mn@doi [\aap] {10.1051/0004-6361/201016276}, \href
  {http://adsabs.harvard.edu/abs/2011A%26A...530A.138C} {530, A138}

\bibitem[\protect\citeauthoryear{{Claria}, {Piatti}  \& {Osborn}}{{Claria}
  et~al.}{1996}]{claria96}
{Claria} J.~J.,  {Piatti} A.~E.,   {Osborn} W.,  1996, \mn@doi [\pasp]
  {10.1086/133784}, \href {http://adsabs.harvard.edu/abs/1996PASP..108..672C}
  {108, 672}

\bibitem[\protect\citeauthoryear{{Clari{\'a}}, {Piatti}, {Lapasset}  \&
  {Mermilliod}}{{Clari{\'a}} et~al.}{2003}]{claria03}
{Clari{\'a}} J.~J.,  {Piatti} A.~E.,  {Lapasset} E.,   {Mermilliod} J.-C.,
  2003, \mn@doi [\aap] {10.1051/0004-6361:20021828}, \href
  {http://adsabs.harvard.edu/abs/2003A%26A...399..543C} {399, 543}

\bibitem[\protect\citeauthoryear{{Clementini}, {Gratton}, {Carretta}  \&
  {Sneden}}{{Clementini} et~al.}{1999}]{clementini99}
{Clementini} G.,  {Gratton} R.~G.,  {Carretta} E.,   {Sneden} C.,  1999,
  \mn@doi [\mnras] {10.1046/j.1365-8711.1999.02098.x}, \href
  {http://adsabs.harvard.edu/abs/1999MNRAS.302...22C} {302, 22}

\bibitem[\protect\citeauthoryear{{Cramer}}{{Cramer}}{1999}]{cramer99}
{Cramer} N.,  1999, \mn@doi [\nar] {10.1016/S1387-6473(99)00041-X}, \href
  {http://adsabs.harvard.edu/abs/1999NewAR..43..343C} {43, 343}

\bibitem[\protect\citeauthoryear{{Crawford}}{{Crawford}}{1975}]{crawford75}
{Crawford} D.~L.,  1975, \mn@doi [\aj] {10.1086/111828}, \href
  {http://adsabs.harvard.edu/abs/1975AJ.....80..955C} {80, 955}

\bibitem[\protect\citeauthoryear{{Drazdauskas}, {Tautvai{\v s}ien{\.e}},
  {Smiljanic}, {Bagdonas}  \& {Chorniy}}{{Drazdauskas}
  et~al.}{2016}]{drazdauskas16}
{Drazdauskas} A.,  {Tautvai{\v s}ien{\.e}} G.,  {Smiljanic} R.,  {Bagdonas} V.,
    {Chorniy} Y.,  2016, \mn@doi [\mnras] {10.1093/mnras/stw1701}, \href
  {http://adsabs.harvard.edu/abs/2016MNRAS.462..794D} {462, 794}

\bibitem[\protect\citeauthoryear{{Fulbright}}{{Fulbright}}{2000}]{fulbright00}
{Fulbright} J.~P.,  2000, \mn@doi [\aj] {10.1086/301548}, \href
  {http://adsabs.harvard.edu/abs/2000AJ....120.1841F} {120, 1841}

\bibitem[\protect\citeauthoryear{{Gaia Collaboration} et~al.,}{{Gaia
  Collaboration} et~al.}{2016}]{gaia2016}
{Gaia Collaboration} et~al., 2016, \mn@doi [\aap]
  {10.1051/0004-6361/201629512}, \href
  {http://adsabs.harvard.edu/abs/2016A%26A...595A...2G} {595, A2}

\bibitem[\protect\citeauthoryear{{Gebran}, {Vick}, {Monier}  \&
  {Fossati}}{{Gebran} et~al.}{2010}]{gebran10}
{Gebran} M.,  {Vick} M.,  {Monier} R.,   {Fossati} L.,  2010, \mn@doi [\aap]
  {10.1051/0004-6361/200913273}, \href
  {http://adsabs.harvard.edu/abs/2010A%26A...523A..71G} {523, A71}

\bibitem[\protect\citeauthoryear{{Gilmore} et~al.,}{{Gilmore}
  et~al.}{2012}]{gilmore12}
{Gilmore} G.,  et~al., 2012, The Messenger, \href
  {http://adsabs.harvard.edu/abs/2012Msngr.147...25G} {147, 25}

\bibitem[\protect\citeauthoryear{{Golay}}{{Golay}}{1972}]{golay72}
{Golay} M.,  1972, \mn@doi [Vistas in Astronomy]
  {10.1016/0083-6656(72)90023-2}, \href
  {http://adsabs.harvard.edu/abs/1972VA.....14...13G} {14, 13}

\bibitem[\protect\citeauthoryear{{Golay}}{{Golay}}{1980}]{golay80}
{Golay} M.,  1980, \mn@doi [Vistas in Astronomy]
  {10.1016/0083-6656(80)90028-8}, \href
  {http://adsabs.harvard.edu/abs/1980VA.....24..141G} {24, 141}

\bibitem[\protect\citeauthoryear{{Grenon}}{{Grenon}}{1978}]{grenon78}
{Grenon} M.,  1978, {Photometric properties of G, K and M stars in relation to
  galactic structure and evolution}.
Publ. Obs. Gen\`{e}ve, S\'{e}r. B, Fasc. 5

\bibitem[\protect\citeauthoryear{{Grenon}}{{Grenon}}{1981}]{grenon81}
{Grenon} M.,  1981, in {Philip} A.~G.~D.,  {Hayes} D.~S.,  eds,  Proceedings of
  IAU Colloquium Vol. 68, Astrophysical Parameters for Globular Clusters.
  p.~393

\bibitem[\protect\citeauthoryear{{Hauck}}{{Hauck}}{1973}]{hauck73}
{Hauck} B.,  1973, in {Hauck} B.,  {Westerlund} B.~E.,  eds,  IAU Symposium
  Vol. 54, Problems of Calibration of Absolute Magnitudes and Temperature of
  Stars. p.~117

\bibitem[\protect\citeauthoryear{{Hauck}}{{Hauck}}{1994}]{hauck94}
{Hauck} B.,  1994, in {Corbally} C.~J.,  {Gray} R.~O.,   {Garrison} R.~F.,
  eds,  Astronomical Society of the Pacific Conference Series Vol. 60, The MK
  Process at 50 Years: A Powerful Tool for Astrophysical Insight. p.~157

\bibitem[\protect\citeauthoryear{{Hauck} \& {North}}{{Hauck} \&
  {North}}{1982}]{hauck82}
{Hauck} B.,  {North} P.,  1982, \aap, \href
  {http://adsabs.harvard.edu/abs/1982A%26A...114...23H} {114, 23}

\bibitem[\protect\citeauthoryear{{Hauck}, {Jaschek}, {Jaschek}  \&
  {Andrillat}}{{Hauck} et~al.}{1991}]{hauck91}
{Hauck} B.,  {Jaschek} C.,  {Jaschek} M.,   {Andrillat} Y.,  1991, \aap, \href
  {http://adsabs.harvard.edu/abs/1991A%26A...252..260H} {252, 260}

\bibitem[\protect\citeauthoryear{{Heiter}, {Soubiran}, {Netopil}  \&
  {Paunzen}}{{Heiter} et~al.}{2014}]{heiter14}
{Heiter} U.,  {Soubiran} C.,  {Netopil} M.,   {Paunzen} E.,  2014, \mn@doi
  [\aap] {10.1051/0004-6361/201322559}, \href
  {http://adsabs.harvard.edu/abs/2014A%26A...561A..93H} {561, A93}

\bibitem[\protect\citeauthoryear{{Heiter}, {Jofr{\'e}}, {Gustafsson}, {Korn},
  {Soubiran}  \& {Th{\'e}venin}}{{Heiter} et~al.}{2015}]{heiter15}
{Heiter} U.,  {Jofr{\'e}} P.,  {Gustafsson} B.,  {Korn} A.~J.,  {Soubiran} C.,
   {Th{\'e}venin} F.,  2015, \mn@doi [\aap] {10.1051/0004-6361/201526319},
  \href {http://adsabs.harvard.edu/abs/2015A%26A...582A..49H} {582, A49}

\bibitem[\protect\citeauthoryear{{Hekker} \& {Mel{\'e}ndez}}{{Hekker} \&
  {Mel{\'e}ndez}}{2007}]{hekker07}
{Hekker} S.,  {Mel{\'e}ndez} J.,  2007, \mn@doi [\aap]
  {10.1051/0004-6361:20078233}, \href
  {http://adsabs.harvard.edu/abs/2007A%26A...475.1003H} {475, 1003}

\bibitem[\protect\citeauthoryear{{Huang}, {Liu}, {Yuan}, {Xiang}, {Chen}  \&
  {Zhang}}{{Huang} et~al.}{2015}]{huang15}
{Huang} Y.,  {Liu} X.-W.,  {Yuan} H.-B.,  {Xiang} M.-S.,  {Chen} B.-Q.,
  {Zhang} H.-W.,  2015, \mn@doi [\mnras] {10.1093/mnras/stv1991}, \href
  {http://adsabs.harvard.edu/abs/2015MNRAS.454.2863H} {454, 2863}

\bibitem[\protect\citeauthoryear{{Isaacson} \& {Fischer}}{{Isaacson} \&
  {Fischer}}{2010}]{isaacson10}
{Isaacson} H.,  {Fischer} D.,  2010, \mn@doi [\apj]
  {10.1088/0004-637X/725/1/875}, \href
  {http://adsabs.harvard.edu/abs/2010ApJ...725..875I} {725, 875}

\bibitem[\protect\citeauthoryear{{Jofr{\'e}} et~al.,}{{Jofr{\'e}}
  et~al.}{2014}]{jofre14}
{Jofr{\'e}} P.,  et~al., 2014, \mn@doi [\aap] {10.1051/0004-6361/201322440},
  \href {http://adsabs.harvard.edu/abs/2014A%26A...564A.133J} {564, A133}

\bibitem[\protect\citeauthoryear{{Jofr{\'e}}, {Petrucci}, {Saffe}, {Saker}, {de
  la Villarmois}, {Chavero}, {G{\'o}mez}  \& {Mauas}}{{Jofr{\'e}}
  et~al.}{2015}]{jofre15}
{Jofr{\'e}} E.,  {Petrucci} R.,  {Saffe} C.,  {Saker} L.,  {de la Villarmois}
  E.~A.,  {Chavero} C.,  {G{\'o}mez} M.,   {Mauas} P.~J.~D.,  2015, \mn@doi
  [\aap] {10.1051/0004-6361/201424474}, \href
  {http://adsabs.harvard.edu/abs/2015A%26A...574A..50J} {574, A50}

\bibitem[\protect\citeauthoryear{{Jones}, {Robinson}, {Slee}  \&
  {Stewart}}{{Jones} et~al.}{1992}]{jones92}
{Jones} K.~L.,  {Robinson} R.~D.,  {Slee} O.~B.,   {Stewart} R.~T.,  1992,
  \mn@doi [\mnras] {10.1093/mnras/256.3.535}, \href
  {http://adsabs.harvard.edu/abs/1992MNRAS.256..535J} {256, 535}

\bibitem[\protect\citeauthoryear{{Jones}, {Jenkins}, {Rojo}  \& {Melo}}{{Jones}
  et~al.}{2011}]{jones11}
{Jones} M.~I.,  {Jenkins} J.~S.,  {Rojo} P.,   {Melo} C.~H.~F.,  2011, \mn@doi
  [\aap] {10.1051/0004-6361/201117887}, \href
  {http://adsabs.harvard.edu/abs/2011A%26A...536A..71J} {536, A71}

\bibitem[\protect\citeauthoryear{{Karaali}, {Bilir}, {Ak}, {Yaz}  \& {Co{\c
  s}kuno{\u g}lu}}{{Karaali} et~al.}{2011}]{karaali11}
{Karaali} S.,  {Bilir} S.,  {Ak} S.,  {Yaz} E.,   {Co{\c s}kuno{\u g}lu} B.,
  2011, \mn@doi [\pasa] {10.1071/AS10026}, \href
  {http://adsabs.harvard.edu/abs/2011PASA...28...95K} {28, 95}

\bibitem[\protect\citeauthoryear{{Karata{\c s}} \& {Schuster}}{{Karata{\c s}}
  \& {Schuster}}{2006}]{karatas06}
{Karata{\c s}} Y.,  {Schuster} W.~J.,  2006, \mn@doi [\mnras]
  {10.1111/j.1365-2966.2006.10800.x}, \href
  {http://adsabs.harvard.edu/abs/2006MNRAS.371.1793K} {371, 1793}

\bibitem[\protect\citeauthoryear{{Karata{\c s}} \& {Schuster}}{{Karata{\c s}}
  \& {Schuster}}{2010}]{karatas10}
{Karata{\c s}} Y.,  {Schuster} W.~J.,  2010, \mn@doi [\na]
  {10.1016/j.newast.2009.12.003}, \href
  {http://adsabs.harvard.edu/abs/2010NewA...15..444K} {15, 444}

\bibitem[\protect\citeauthoryear{{K{\i}l{\i}{\c c}o{\u g}lu}, {Monier},
  {Richer}, {Fossati}  \& {Albayrak}}{{K{\i}l{\i}{\c c}o{\u g}lu}
  et~al.}{2016}]{kilicoglu16}
{K{\i}l{\i}{\c c}o{\u g}lu} T.,  {Monier} R.,  {Richer} J.,  {Fossati} L.,
  {Albayrak} B.,  2016, \mn@doi [\aj] {10.3847/0004-6256/151/3/49}, \href
  {http://adsabs.harvard.edu/abs/2016AJ....151...49K} {151, 49}

\bibitem[\protect\citeauthoryear{{Kunzli}, {North}, {Kurucz}  \&
  {Nicolet}}{{Kunzli} et~al.}{1997}]{kunzli97}
{Kunzli} M.,  {North} P.,  {Kurucz} R.~L.,   {Nicolet} B.,  1997, \mn@doi
  [\aaps] {10.1051/aas:1997291}, \href
  {http://adsabs.harvard.edu/abs/1997A%26AS..122...51K} {122}

\bibitem[\protect\citeauthoryear{{Langer} \& {Kudritzki}}{{Langer} \&
  {Kudritzki}}{2014}]{langer14}
{Langer} N.,  {Kudritzki} R.~P.,  2014, \mn@doi [\aap]
  {10.1051/0004-6361/201423374}, \href
  {http://adsabs.harvard.edu/abs/2014A%26A...564A..52L} {564, A52}

\bibitem[\protect\citeauthoryear{{Lapasset}, {Clari{\'a}}  \&
  {Mermilliod}}{{Lapasset} et~al.}{2000}]{lapa00}
{Lapasset} E.,  {Clari{\'a}} J.~J.,   {Mermilliod} J.-C.,  2000, \aap, \href
  {http://adsabs.harvard.edu/abs/2000A%26A...361..945L} {361, 945}

\bibitem[\protect\citeauthoryear{{Liu}, {Yong}, {Asplund}, {Ram{\'{\i}}rez}  \&
  {Mel{\'e}ndez}}{{Liu} et~al.}{2016}]{liu16}
{Liu} F.,  {Yong} D.,  {Asplund} M.,  {Ram{\'{\i}}rez} I.,   {Mel{\'e}ndez} J.,
   2016, \mn@doi [\mnras] {10.1093/mnras/stw247}, \href
  {http://adsabs.harvard.edu/abs/2016MNRAS.457.3934L} {457, 3934}

\bibitem[\protect\citeauthoryear{{Luck}}{{Luck}}{2014}]{luck14}
{Luck} R.~E.,  2014, \mn@doi [\aj] {10.1088/0004-6256/147/6/137}, \href
  {http://adsabs.harvard.edu/abs/2014AJ....147..137L} {147, 137}

\bibitem[\protect\citeauthoryear{{Luck} \& {Heiter}}{{Luck} \&
  {Heiter}}{2005}]{luck05}
{Luck} R.~E.,  {Heiter} U.,  2005, \mn@doi [\aj] {10.1086/427250}, \href
  {http://adsabs.harvard.edu/abs/2005AJ....129.1063L} {129, 1063}

\bibitem[\protect\citeauthoryear{{Luck} \& {Heiter}}{{Luck} \&
  {Heiter}}{2006}]{luck06}
{Luck} R.~E.,  {Heiter} U.,  2006, \mn@doi [\aj] {10.1086/504080}, \href
  {http://adsabs.harvard.edu/abs/2006AJ....131.3069L} {131, 3069}

\bibitem[\protect\citeauthoryear{{Lynga} \& {Wramdemark}}{{Lynga} \&
  {Wramdemark}}{1984}]{lynga84}
{Lynga} G.,  {Wramdemark} S.,  1984, \aap, \href
  {http://adsabs.harvard.edu/abs/1984A%26A...132...58L} {132, 58}

\bibitem[\protect\citeauthoryear{{Mallik}}{{Mallik}}{1998}]{mallik98}
{Mallik} S.~V.,  1998, \aap, \href
  {http://adsabs.harvard.edu/abs/1998A%26A...338..623M} {338, 623}

\bibitem[\protect\citeauthoryear{{McWilliam}}{{McWilliam}}{1990}]{mcwilliam90}
{McWilliam} A.,  1990, \mn@doi [\apjs] {10.1086/191527}, \href
  {http://adsabs.harvard.edu/abs/1990ApJS...74.1075M} {74, 1075}

\bibitem[\protect\citeauthoryear{{Mel{\'e}ndez} \&
  {Ram{\'{\i}}rez}}{{Mel{\'e}ndez} \& {Ram{\'{\i}}rez}}{2003}]{melendez03}
{Mel{\'e}ndez} J.,  {Ram{\'{\i}}rez} I.,  2003, \mn@doi [\aap]
  {10.1051/0004-6361:20021681}, \href
  {http://adsabs.harvard.edu/abs/2003A%26A...398..705M} {398, 705}

\bibitem[\protect\citeauthoryear{{Mermilliod} \& {Paunzen}}{{Mermilliod} \&
  {Paunzen}}{2003}]{mermilliod03}
{Mermilliod} J.-C.,  {Paunzen} E.,  2003, \mn@doi [\aap]
  {10.1051/0004-6361:20031112}, \href
  {http://adsabs.harvard.edu/abs/2003A%26A...410..511M} {410, 511}

\bibitem[\protect\citeauthoryear{{Mermilliod}, {Mermilliod}  \&
  {Hauck}}{{Mermilliod} et~al.}{1997}]{mmh97}
{Mermilliod} J.-C.,  {Mermilliod} M.,   {Hauck} B.,  1997, \mn@doi [\aaps]
  {10.1051/aas:1997197}, \href
  {http://adsabs.harvard.edu/abs/1997A%26AS..124..349M} {124}

\bibitem[\protect\citeauthoryear{{Mermilliod}, {Mayor}  \& {Udry}}{{Mermilliod}
  et~al.}{2008}]{mermilliod08}
{Mermilliod} J.~C.,  {Mayor} M.,   {Udry} S.,  2008, \mn@doi [\aap]
  {10.1051/0004-6361:200809664}, \href
  {http://adsabs.harvard.edu/abs/2008A%26A...485..303M} {485, 303}

\bibitem[\protect\citeauthoryear{{Mishenina}, {Soubiran}, {Bienaym{\'e}},
  {Korotin}, {Belik}, {Usenko}  \& {Kovtyukh}}{{Mishenina}
  et~al.}{2008}]{mishenina08}
{Mishenina} T.~V.,  {Soubiran} C.,  {Bienaym{\'e}} O.,  {Korotin} S.~A.,
  {Belik} S.~I.,  {Usenko} I.~A.,   {Kovtyukh} V.~V.,  2008, \mn@doi [\aap]
  {10.1051/0004-6361:200810360}, \href
  {http://adsabs.harvard.edu/abs/2008A%26A...489..923M} {489, 923}

\bibitem[\protect\citeauthoryear{{Mishenina}, {Soubiran}, {Kovtyukh}, {Katsova}
   \& {Livshits}}{{Mishenina} et~al.}{2012}]{mishenina12}
{Mishenina} T.~V.,  {Soubiran} C.,  {Kovtyukh} V.~V.,  {Katsova} M.~M.,
  {Livshits} M.~A.,  2012, \mn@doi [\aap] {10.1051/0004-6361/201118412}, \href
  {http://adsabs.harvard.edu/abs/2012A%26A...547A.106M} {547, A106}

\bibitem[\protect\citeauthoryear{{Molenda-{\.Z}akowicz}, {Brogaard},
  {Niemczura}, {Bergemann}, {Frasca}, {Arentoft}  \&
  {Grundahl}}{{Molenda-{\.Z}akowicz} et~al.}{2014}]{molenda14}
{Molenda-{\.Z}akowicz} J.,  {Brogaard} K.,  {Niemczura} E.,  {Bergemann} M.,
  {Frasca} A.,  {Arentoft} T.,   {Grundahl} F.,  2014, \mn@doi [\mnras]
  {10.1093/mnras/stu1934}, \href
  {http://adsabs.harvard.edu/abs/2014MNRAS.445.2446M} {445, 2446}

\bibitem[\protect\citeauthoryear{{Morel} et~al.,}{{Morel}
  et~al.}{2014}]{morel14}
{Morel} T.,  et~al., 2014, \mn@doi [\aap] {10.1051/0004-6361/201322810}, \href
  {http://adsabs.harvard.edu/abs/2014A%26A...564A.119M} {564, A119}

\bibitem[\protect\citeauthoryear{{Mowlavi}, {Barblan}, {Saesen}  \&
  {Eyer}}{{Mowlavi} et~al.}{2013}]{mowlavi13}
{Mowlavi} N.,  {Barblan} F.,  {Saesen} S.,   {Eyer} L.,  2013, \mn@doi [\aap]
  {10.1051/0004-6361/201321065}, \href
  {http://adsabs.harvard.edu/abs/2013A%26A...554A.108M} {554, A108}

\bibitem[\protect\citeauthoryear{{Murphy} et~al.,}{{Murphy}
  et~al.}{2015}]{murphy15}
{Murphy} S.~J.,  et~al., 2015, \mn@doi [\pasa] {10.1017/pasa.2015.34}, \href
  {http://adsabs.harvard.edu/abs/2015PASA...32...36M} {32, e036}

\bibitem[\protect\citeauthoryear{{Napiwotzki}, {Schoenberner}  \&
  {Wenske}}{{Napiwotzki} et~al.}{1993}]{napiwotzki93}
{Napiwotzki} R.,  {Schoenberner} D.,   {Wenske} V.,  1993, \aap, \href
  {http://adsabs.harvard.edu/abs/1993A%26A...268..653N} {268, 653}

\bibitem[\protect\citeauthoryear{{Netopil} \& {Paunzen}}{{Netopil} \&
  {Paunzen}}{2013}]{netopil13}
{Netopil} M.,  {Paunzen} E.,  2013, \mn@doi [\aap]
  {10.1051/0004-6361/201321829}, \href
  {http://adsabs.harvard.edu/abs/2013A%26A...557A..10N} {557, A10}

\bibitem[\protect\citeauthoryear{{Netopil}, {Paunzen}, {Heiter}  \&
  {Soubiran}}{{Netopil} et~al.}{2016}]{netopil16}
{Netopil} M.,  {Paunzen} E.,  {Heiter} U.,   {Soubiran} C.,  2016, \mn@doi
  [\aap] {10.1051/0004-6361/201526370}, \href
  {http://adsabs.harvard.edu/abs/2016A%26A...585A.150N} {585, A150}

\bibitem[\protect\citeauthoryear{{Pasquini}, {Liu}  \&
  {Pallavicini}}{{Pasquini} et~al.}{1994}]{pasquini94}
{Pasquini} L.,  {Liu} Q.,   {Pallavicini} R.,  1994, \aap, \href
  {http://adsabs.harvard.edu/abs/1994A%26A...287..191P} {287, 191}

\bibitem[\protect\citeauthoryear{{Paulson} \& {Yelda}}{{Paulson} \&
  {Yelda}}{2006}]{paulson06}
{Paulson} D.~B.,  {Yelda} S.,  2006, \mn@doi [\pasp] {10.1086/504115}, \href
  {http://adsabs.harvard.edu/abs/2006PASP..118..706P} {118, 706}

\bibitem[\protect\citeauthoryear{{Paunzen}}{{Paunzen}}{2015}]{paunzen15}
{Paunzen} E.,  2015, \mn@doi [\aap] {10.1051/0004-6361/201526413}, \href
  {http://adsabs.harvard.edu/abs/2015A%26A...580A..23P} {580, A23}

\bibitem[\protect\citeauthoryear{{Paunzen}, {Heiter}, {Netopil}  \&
  {Soubiran}}{{Paunzen} et~al.}{2010}]{paunzen10}
{Paunzen} E.,  {Heiter} U.,  {Netopil} M.,   {Soubiran} C.,  2010, \mn@doi
  [\aap] {10.1051/0004-6361/201014131}, \href
  {http://adsabs.harvard.edu/abs/2010A%26A...517A..32P} {517, A32}

\bibitem[\protect\citeauthoryear{{Perryman} et~al.,}{{Perryman}
  et~al.}{1998}]{perryman98}
{Perryman} M.~A.~C.,  et~al., 1998, \aap, \href
  {http://cdsads.u-strasbg.fr/abs/1998A%26A...331...81P} {331, 81}

\bibitem[\protect\citeauthoryear{{Pomp{\'e}ia} et~al.,}{{Pomp{\'e}ia}
  et~al.}{2011}]{pompeia11}
{Pomp{\'e}ia} L.,  et~al., 2011, \mn@doi [\mnras]
  {10.1111/j.1365-2966.2011.18685.x}, \href
  {http://adsabs.harvard.edu/abs/2011MNRAS.415.1138P} {415, 1138}

\bibitem[\protect\citeauthoryear{{Qui}, {Zhao}, {Takada-Hidai}, {Chen},
  {Takeda}, {Noguchi}, {Sadakane}  \& {Aoki}}{{Qui} et~al.}{2002}]{qui02}
{Qui} H.-M.,  {Zhao} G.,  {Takada-Hidai} M.,  {Chen} Y.-Q.,  {Takeda} Y.,
  {Noguchi} K.,  {Sadakane} K.,   {Aoki} W.,  2002, \mn@doi [\pasj]
  {10.1093/pasj/54.1.103}, \href
  {http://adsabs.harvard.edu/abs/2002PASJ...54..103Q} {54, 103}

\bibitem[\protect\citeauthoryear{{Raboud}, {Cramer}  \& {Bernasconi}}{{Raboud}
  et~al.}{1997}]{raboud97}
{Raboud} D.,  {Cramer} N.,   {Bernasconi} P.~A.,  1997, \aap, \href
  {http://adsabs.harvard.edu/abs/1997A%26A...325..167R} {325, 167}

\bibitem[\protect\citeauthoryear{{Ram{\'{\i}}rez}, {Allende Prieto}  \&
  {Lambert}}{{Ram{\'{\i}}rez} et~al.}{2013}]{ramirez13}
{Ram{\'{\i}}rez} I.,  {Allende Prieto} C.,   {Lambert} D.~L.,  2013, \mn@doi
  [\apj] {10.1088/0004-637X/764/1/78}, \href
  {http://adsabs.harvard.edu/abs/2013ApJ...764...78R} {764, 78}

\bibitem[\protect\citeauthoryear{{Randich}, {Gratton}, {Pallavicini},
  {Pasquini}  \& {Carretta}}{{Randich} et~al.}{1999}]{randich99}
{Randich} S.,  {Gratton} R.,  {Pallavicini} R.,  {Pasquini} L.,   {Carretta}
  E.,  1999, \aap, \href {http://adsabs.harvard.edu/abs/1999A%26A...348..487R}
  {348, 487}

\bibitem[\protect\citeauthoryear{{Reddy}, {Giridhar}  \& {Lambert}}{{Reddy}
  et~al.}{2013}]{reddy13}
{Reddy} A.~B.~S.,  {Giridhar} S.,   {Lambert} D.~L.,  2013, \mn@doi [\mnras]
  {10.1093/mnras/stt412}, \href
  {http://adsabs.harvard.edu/abs/2013MNRAS.431.3338R} {431, 3338}

\bibitem[\protect\citeauthoryear{{Reddy}, {Giridhar}  \& {Lambert}}{{Reddy}
  et~al.}{2015}]{reddy15}
{Reddy} A.~B.~S.,  {Giridhar} S.,   {Lambert} D.~L.,  2015, \mn@doi [\mnras]
  {10.1093/mnras/stv908}, \href
  {http://adsabs.harvard.edu/abs/2015MNRAS.450.4301R} {450, 4301}

\bibitem[\protect\citeauthoryear{{Renson} \& {Manfroid}}{{Renson} \&
  {Manfroid}}{2009}]{renson09}
{Renson} P.,  {Manfroid} J.,  2009, \mn@doi [\aap]
  {10.1051/0004-6361/200810788}, \href
  {http://adsabs.harvard.edu/abs/2009A%26A...498..961R} {498, 961}

\bibitem[\protect\citeauthoryear{{Roederer}, {Preston}, {Thompson}, {Shectman},
  {Sneden}, {Burley}  \& {Kelson}}{{Roederer} et~al.}{2014}]{roederer14}
{Roederer} I.~U.,  {Preston} G.~W.,  {Thompson} I.~B.,  {Shectman} S.~A.,
  {Sneden} C.,  {Burley} G.~S.,   {Kelson} D.~D.,  2014, \mn@doi [\aj]
  {10.1088/0004-6256/147/6/136}, \href
  {http://adsabs.harvard.edu/abs/2014AJ....147..136R} {147, 136}

\bibitem[\protect\citeauthoryear{{Santos}, {Lovis}, {Pace}, {Melendez}  \&
  {Naef}}{{Santos} et~al.}{2009}]{santos09}
{Santos} N.~C.,  {Lovis} C.,  {Pace} G.,  {Melendez} J.,   {Naef} D.,  2009,
  \mn@doi [\aap] {10.1051/0004-6361:200811093}, \href
  {http://cdsads.u-strasbg.fr/abs/2009A%26A...493..309S} {493, 309}

\bibitem[\protect\citeauthoryear{{Skiff}}{{Skiff}}{2014}]{skiff14}
{Skiff} B.~A.,  2014, VizieR Online Data Catalog, \href
  {http://cdsads.u-strasbg.fr/abs/2014yCat....1.2023S} {1, 2023}

\bibitem[\protect\citeauthoryear{{Soubiran}, {Bienaym{\'e}}  \&
  {Siebert}}{{Soubiran} et~al.}{2003}]{soubiran03}
{Soubiran} C.,  {Bienaym{\'e}} O.,   {Siebert} A.,  2003, \mn@doi [\aap]
  {10.1051/0004-6361:20021615}, \href
  {http://adsabs.harvard.edu/abs/2003A%26A...398..141S} {398, 141}

\bibitem[\protect\citeauthoryear{{Soubiran}, {Le Campion}, {Brouillet}  \&
  {Chemin}}{{Soubiran} et~al.}{2016}]{soubiran16}
{Soubiran} C.,  {Le Campion} J.-F.,  {Brouillet} N.,   {Chemin} L.,  2016,
  \mn@doi [\aap] {10.1051/0004-6361/201628497}, \href
  {http://adsabs.harvard.edu/abs/2016A%26A...591A.118S} {591, A118}

\bibitem[\protect\citeauthoryear{{Sousa} et~al.,}{{Sousa}
  et~al.}{2008}]{sousa08}
{Sousa} S.~G.,  et~al., 2008, \mn@doi [\aap] {10.1051/0004-6361:200809698},
  \href {http://adsabs.harvard.edu/abs/2008A%26A...487..373S} {487, 373}

\bibitem[\protect\citeauthoryear{{Sousa}, {Santos}, {Israelian}, {Mayor}  \&
  {Udry}}{{Sousa} et~al.}{2011}]{sousa11}
{Sousa} S.~G.,  {Santos} N.~C.,  {Israelian} G.,  {Mayor} M.,   {Udry} S.,
  2011, \mn@doi [\aap] {10.1051/0004-6361/201117699}, \href
  {http://adsabs.harvard.edu/abs/2011A%26A...533A.141S} {533, A141}

\bibitem[\protect\citeauthoryear{{Spina} et~al.,}{{Spina}
  et~al.}{2017}]{spina17}
{Spina} L.,  et~al., 2017, preprint, \href
  {http://adsabs.harvard.edu/abs/2017arXiv170203461S} {} (\mn@eprint {arXiv}
  {1702.03461})

\bibitem[\protect\citeauthoryear{{Th{\'e}venin} \& {Idiart}}{{Th{\'e}venin} \&
  {Idiart}}{1999}]{thevenin99}
{Th{\'e}venin} F.,  {Idiart} T.~P.,  1999, \mn@doi [\apj] {10.1086/307578},
  \href {http://adsabs.harvard.edu/abs/1999ApJ...521..753T} {521, 753}

\bibitem[\protect\citeauthoryear{{Tomkin}, {Lemke}, {Lambert}  \&
  {Sneden}}{{Tomkin} et~al.}{1992}]{tomkin92}
{Tomkin} J.,  {Lemke} M.,  {Lambert} D.~L.,   {Sneden} C.,  1992, \mn@doi [\aj]
  {10.1086/116342}, \href {http://adsabs.harvard.edu/abs/1992AJ....104.1568T}
  {104, 1568}

\bibitem[\protect\citeauthoryear{{Trevisan} \& {Barbuy}}{{Trevisan} \&
  {Barbuy}}{2014}]{trevisan14}
{Trevisan} M.,  {Barbuy} B.,  2014, \mn@doi [\aap]
  {10.1051/0004-6361/201322967}, \href
  {http://adsabs.harvard.edu/abs/2014A%26A...570A..22T} {570, A22}

\bibitem[\protect\citeauthoryear{{Trevisan}, {Barbuy}, {Eriksson},
  {Gustafsson}, {Grenon}  \& {Pomp{\'e}ia}}{{Trevisan}
  et~al.}{2011}]{trevisan11}
{Trevisan} M.,  {Barbuy} B.,  {Eriksson} K.,  {Gustafsson} B.,  {Grenon} M.,
  {Pomp{\'e}ia} L.,  2011, \mn@doi [\aap] {10.1051/0004-6361/201016056}, \href
  {http://cdsads.u-strasbg.fr/abs/2011A%26A...535A..42T} {535, A42}

\bibitem[\protect\citeauthoryear{{Tsantaki}, {Sousa}, {Adibekyan}, {Santos},
  {Mortier}  \& {Israelian}}{{Tsantaki} et~al.}{2013}]{tsantaki13}
{Tsantaki} M.,  {Sousa} S.~G.,  {Adibekyan} V.~Z.,  {Santos} N.~C.,  {Mortier}
  A.,   {Israelian} G.,  2013, \mn@doi [\aap] {10.1051/0004-6361/201321103},
  \href {http://adsabs.harvard.edu/abs/2013A%26A...555A.150T} {555, A150}

\bibitem[\protect\citeauthoryear{{Twarog}, {Ashman}  \&
  {Anthony-Twarog}}{{Twarog} et~al.}{1997}]{twarog97}
{Twarog} B.~A.,  {Ashman} K.~M.,   {Anthony-Twarog} B.~J.,  1997, \mn@doi [\aj]
  {10.1086/118667}, \href {http://adsabs.harvard.edu/abs/1997AJ....114.2556T}
  {114, 2556}

\bibitem[\protect\citeauthoryear{{Valenti} \& {Fischer}}{{Valenti} \&
  {Fischer}}{2005}]{valenti05}
{Valenti} J.~A.,  {Fischer} D.~A.,  2005, \mn@doi [\apjs] {10.1086/430500},
  \href {http://adsabs.harvard.edu/abs/2005ApJS..159..141V} {159, 141}

\bibitem[\protect\citeauthoryear{{Varenne} \& {Monier}}{{Varenne} \&
  {Monier}}{1999}]{varenne99}
{Varenne} O.,  {Monier} R.,  1999, \aap, \href
  {http://adsabs.harvard.edu/abs/1999A%26A...351..247V} {351, 247}

\bibitem[\protect\citeauthoryear{{Zhao} \& {Magain}}{{Zhao} \&
  {Magain}}{1990}]{zhao90}
{Zhao} G.,  {Magain} P.,  1990, \aap, \href
  {http://adsabs.harvard.edu/abs/1990A%26A...238..242Z} {238, 242}

\bibitem[\protect\citeauthoryear{{Zhao}, {Qiu}  \& {Mao}}{{Zhao}
  et~al.}{2001}]{zhao01}
{Zhao} G.,  {Qiu} H.~M.,   {Mao} S.,  2001, \mn@doi [\apjl] {10.1086/319832},
  \href {http://adsabs.harvard.edu/abs/2001ApJ...551L..85Z} {551, L85}

\bibitem[\protect\citeauthoryear{{da Silva} et~al.,}{{da Silva}
  et~al.}{2006}]{dasilva06}
{da Silva} L.,  et~al., 2006, \mn@doi [\aap] {10.1051/0004-6361:20065105},
  \href {http://adsabs.harvard.edu/abs/2006A%26A...458..609D} {458, 609}

\bibitem[\protect\citeauthoryear{{da Silva}, {Milone}  \& {Reddy}}{{da Silva}
  et~al.}{2011}]{dasilva11}
{da Silva} R.,  {Milone} A.~C.,   {Reddy} B.~E.,  2011, \mn@doi [\aap]
  {10.1051/0004-6361/201015907}, \href
  {http://adsabs.harvard.edu/abs/2011A%26A...526A..71D} {526, A71}

\bibitem[\protect\citeauthoryear{{da Silva}, {Milone}  \& {Rocha-Pinto}}{{da
  Silva} et~al.}{2015}]{dasilva15}
{da Silva} R.,  {Milone} A.~d.~C.,   {Rocha-Pinto} H.~J.,  2015, \mn@doi [\aap]
  {10.1051/0004-6361/201525770}, \href
  {http://adsabs.harvard.edu/abs/2015A%26A...580A..24D} {580, A24}

\bibitem[\protect\citeauthoryear{{van Leeuwen}}{{van
  Leeuwen}}{2007}]{leeuwen07}
{van Leeuwen} F.,  2007, \mn@doi [\aap] {10.1051/0004-6361:20078357}, \href
  {http://adsabs.harvard.edu/abs/2007A%26A...474..653V} {474, 653}

\makeatother
\end{thebibliography}




\appendix

\section{Some extra material}
\onecolumn
\begin{longtable}{lcclcclcccl}
\caption{\label{tab:metaldata} Metallicities of the open clusters}\\
\hline\hline
Cluster & Calib  & [Fe/H]$_{Geneva}$ & Stars & $E(B-V)$ & Ref & [Fe/H]$_{spec}$ & Stars & [Fe/H]$_{phot}$ & W & No\\ 
        &       & [dex]  &       & [mag]    &     &  [dex]          &       &                 &   &  \\
\hline
\endfirsthead
\caption{Continued.}\\
\hline\hline
Cluster & Calib  & [Fe/H]$_{Geneva}$ & Stars & $E(B-V)$ & Ref & [Fe/H]$_{spec}$ & Stars & [Fe/H]$_{phot}$ & W & No\\ 
        &       & [dex]  &       & [mag]    &     &  [dex]          &       &                 &   &  \\
\hline
\endhead
\hline
\endfoot
\hline
\endlastfoot
Blanco 1 & D & $-$0.03 $\pm$ 0.09 & 20 & 0.03 $\pm$ 0.01 & G & +0.03 $\pm$ 0.07 & 6 & $-$0.03    & 1 & 1 \\
IC 2391 & D & +0.02 $\pm$ 0.09 & 23 & 0.02 $\pm$ 0.01 & G & $-$0.01 $\pm$ 0.03 & 12 & $-$0.03 $\pm$ 0.06 & 2.5 & 2 \\
IC 2488 & G & $-$0.02 $\pm$ 0.05 & 3 & 0.29       & [1]    &   &     & +0.03 $\pm$ 0.07 & 1 & 2 \\
IC 2602 & D & +0.00 $\pm$ 0.09 & 13 & 0.04 $\pm$ 0.02 & G & $-$0.02 $\pm$ 0.02 & 7 & $-$0.02 $\pm$ 0.02 & 2.5 & 2 \\
IC 2714 & G & $-$0.01 $\pm$ 0.05 & 5 & 0.35     & T97 & +0.02 $\pm$ 0.06 & 4 & $-$0.01 $\pm$ 0.00 & 2 & 2 \\
IC 4651 & G & +0.06 $\pm$ 0.04 & 8 & 0.11     & T97 & +0.12 $\pm$ 0.04 & 18 & +0.08 $\pm$ 0.02 & 3 & 3 \\
IC 4665 & D & $-$0.03 $\pm$ 0.11 & 6 & 0.19 $\pm$ 0.03 & G & $-$0.03 $\pm$ 0.04 & 18 & $-$0.05 $\pm$ 0.08 & 3.5 & 3 \\
IC 4756 & D & +0.01 $\pm$ 0.12 & 20 & 0.23     & T97 & +0.02 $\pm$ 0.04 & 15 & $-$0.02 $\pm$ 0.02 & 3 & 3 \\
IC 4756 & G & +0.00 $\pm$ 0.08 & 10 & 0.22 $\pm$ 0.02 & uvby  &      &    &      &    &    \\
Melotte 020 & D & $-$0.02 $\pm$ 0.10 & 47 & 0.11 $\pm$ 0.02 & G & +0.14 $\pm$ 0.11 & 2 & +0.01 $\pm$ 0.03 & 3.5 & 3 \\
Melotte 022 & D & $-$0.01 $\pm$ 0.08 & 52 & 0.05 $\pm$ 0.02 & G & $-$0.01 $\pm$ 0.05 & 10 & $-$0.03 $\pm$ 0.08 & 3.5 & 3 \\
Melotte 025 & G & +0.12 $\pm$ 0.05 & 4 & 0.00     & T97   & +0.13 $\pm$ 0.05 & 61 & +0.15 $\pm$ 0.03 & 3.5 & 4 \\
Melotte 111 & D & $-$0.04 $\pm$ 0.08 & 24 & 0.00     & NP13 & +0.00 $\pm$ 0.08 & 10 & $-$0.04 $\pm$ 0.03 & 3.5 & 3 \\
NGC 0752 & D & $-$0.06 $\pm$ 0.08 & 32 & 0.04     & T97 & $-$0.03 $\pm$ 0.06 & 23 & $-$0.04 $\pm$ 0.08 & 6.5 & 6 \\
NGC 0752 & G & $-$0.07 $\pm$ 0.03 & 14 & 0.04     & T97 &     &    &      &    &    \\
NGC 1039 & D & +0.00 $\pm$ 0.09 & 8 & 0.09 $\pm$ 0.01 & G & +0.02 $\pm$ 0.06 & 7 & +0.01 $\pm$ 0.01 & 2.5 & 2 \\
NGC 1545 & G & $-$0.02 & 1 & 0.38     & [2]   & $-$0.06    & 1 & $-$0.04 $\pm$ 0.03 & 1 & 2 \\
NGC 1662 & D & $-$0.07 $\pm$ 0.11 & 6 & 0.38 $\pm$ 0.04 & G & $-$0.11 $\pm$ 0.01 & 2 & $-$0.02 $\pm$ 0.07 & 3 & 3 \\
NGC 1662 & G & $-$0.06 $\pm$ 0.08 & 2 &         & G &      &    &      &    &    \\
NGC 1912 & G & $-$0.11 $\pm$ 0.03 & 2 & 0.26 $\pm$ 0.01 & ST [3] & $-$0.10 $\pm$ 0.14 & 2 & $-$0.11 & 0.5 & 1 \\
NGC 2168 & G & $-$0.16 & 1 & 0.27 $\pm$ 0.02 & G & $-$0.21 $\pm$ 0.10 $^{a}$    &  9  & $-$0.17 $\pm$ 0.01 & 2.5 & 3 \\
NGC 2251 & G & $-$0.10 $\pm$ 0.16 & 3 & 0.25 $\pm$ 0.02   & ST [4] & $-$0.09    & 1 & $-$0.09 $\pm$ 0.01 & 1 & 2 \\
NGC 2281 & D & $-$0.03 $\pm$ 0.10 & 27 & 0.09 $\pm$ 0.01 & G &     &    & $-$0.02 & 1 & 1 \\
NGC 2281 & G & $-$0.01 $\pm$ 0.01 & 2 &         & G &      &    &     &    &    \\
NGC 2287 & D & $-$0.09 $\pm$ 0.10 & 12 & 0.05 $\pm$ 0.02 & G & $-$0.11 $\pm$ 0.01 & 2 & $-$0.11 $\pm$ 0.07 & 4 & 4 \\
NGC 2287 & G & $-$0.11 $\pm$ 0.09 & 5 &         & G &      &    &      &    &    \\
NGC 2301 & D & +0.04 $\pm$ 0.05 & 6 & 0.06 $\pm$ 0.01 & G &      &    & +0.04 $\pm$ 0.01 & 2.5 & 3 \\
NGC 2360 & G & $-$0.06 $\pm$ 0.05 & 11 & 0.10    & T97 & $-$0.03 $\pm$ 0.06 & 9 & $-$0.05 $\pm$ 0.01 & 2.5 & 3 \\
NGC 2422 & D & +0.14 $\pm$ 0.11 & 13 & 0.10 $\pm$ 0.01 & G &      &    & +0.10 $\pm$ 0.04 & 3.5 & 3 \\
NGC 2423 & G & +0.04 $\pm$ 0.05 & 7 & 0.09     & T97 & +0.08 $\pm$ 0.05 & 3 & +0.06 $\pm$ 0.03 & 3.5 & 3 \\
NGC 2447 & D & $-$0.06 $\pm$ 0.07 & 6 & 0.01     & NP13 & $-$0.05 $\pm$ 0.01 & 3 & $-$0.03 $\pm$ 0.04 & 3.5 & 3 \\
NGC 2447 & G & $-$0.07 $\pm$ 0.05 & 7 & 0.01     & NP13 &      &    &      &    &    \\
NGC 2451A & D & $-$0.03 $\pm$ 0.13 & 5 & 0.03 $\pm$ 0.02 & G &  $-$0.04 $\pm$ 0.04   &  4  & $-$0.06 $\pm$ 0.03 & 2.5 & 2 \\
NGC 2451B & D & +0.08 $\pm$ 0.09 & 3 & 0.13 $\pm$ 0.02 & G &  +0.00 $\pm$ 0.02    &  9  & +0.02 $\pm$ 0.05 & 2 & 2 \\
NGC 2516 & D & +0.01 $\pm$ 0.11 & 35 & 0.14 $\pm$ 0.02 & G & +0.05 $\pm$ 0.11 & 2 & +0.03 $\pm$ 0.03 & 4 & 4 \\
NGC 2539 & G & $-$0.01 $\pm$ 0.04 & 9 & 0.05     & [5]   & $-$0.02 $\pm$ 0.08 & 4 & +0.07 $\pm$ 0.11 & 2 & 2 \\
NGC 2547 & D & +0.00 $\pm$ 0.09 & 10 & 0.07 $\pm$ 0.03 & G &  $-$0.02 $\pm$ 0.02   & 10   & $-$0.05 $\pm$ 0.06 & 2.5 & 2 \\
NGC 2548 & D & $-$0.07 $\pm$ 0.11 & 8 & 0.07 $\pm$ 0.02 & G &     &    & $-$0.02 $\pm$ 0.10 & 1.5 & 2 \\
NGC 2632 & D & +0.16 $\pm$ 0.06 & 92 & 0.01     & NP13 & +0.16 $\pm$ 0.08 & 22 & +0.14 $\pm$ 0.02 & 4 & 4 \\
NGC 2632 & G & +0.14 $\pm$ 0.08 & 3 & 0.01     & NP13 &      &    &      &    &    \\
NGC 2682 & G & +0.00 $\pm$ 0.05 & 16 & 0.04     & T97 & +0.03 $\pm$ 0.05 & 27 & $-$0.01 $\pm$ 0.07 & 5 & 5 \\
NGC 3114 & G & +0.05 $\pm$ 0.01 & 3 & 0.10 $\pm$ 0.02 & G & +0.05 $\pm$ 0.06 & 2 & +0.04 $\pm$ 0.02 & 3 & 3 \\
NGC 3532 & D & +0.02 $\pm$ 0.12 & 49 & 0.06 $\pm$ 0.02 & G & +0.00 $\pm$ 0.07 & 4 & +0.00 $\pm$ 0.02 & 3.5 & 3 \\
NGC 3532 & G & +0.00 $\pm$ 0.06 & 6 &        & G &      &    &      &    &    \\
NGC 3680 & G & $-$0.01 $\pm$ 0.05 & 7 & 0.05     & T97 & $-$0.01 $\pm$ 0.06 & 10 & $-$0.07 $\pm$ 0.09 & 5 & 5 \\
NGC 4349 & G & $-$0.04 $\pm$ 0.09 & 7 & 0.40 $\pm$ 0.04     & ST [6] & $-$0.07 $\pm$ 0.06 & 2 & $-$0.05 $\pm$ 0.01 & 2 & 2 \\
NGC 5316 & G & $-$0.08 $\pm$ 0.10 & 2 & 0.35 $\pm$ 0.02 & G & $-$0.02 $\pm$ 0.05      & 4   & +0.03 $\pm$ 0.15 & 1 & 2 \\
NGC 5460 & D & $-$0.01 $\pm$ 0.15 & 4 & 0.14 $\pm$ 0.02 & G &     &    & +0.02 $\pm$ 0.02 & 2 & 2 \\
NGC 5822 & G & $-$0.01 $\pm$ 0.06 & 12 & 0.14     & T97 & +0.08 $\pm$ 0.08 & 7 & $-$0.02 $\pm$ 0.03 & 4.5 & 5 \\
NGC 6025 & D & $-$0.10 $\pm$ 0.11 & 4 & 0.19 $\pm$ 0.02 & G &      &    & $-$0.10    & 0.5 & 1 \\
NGC 6281 & D & +0.07 $\pm$ 0.11 & 10 & 0.19 $\pm$ 0.03 & G & +0.06 $\pm$ 0.06 & 2 & +0.04 $\pm$ 0.03 & 3 & 3 \\
NGC 6405 & D & +0.05 $\pm$ 0.10 & 16 & 0.17 $\pm$ 0.02 & G & +0.07 $\pm$ 0.03 $^{a}$     & 7   & +0.06 $\pm$ 0.01 & 2.5 & 2 \\
NGC 6475 & D & +0.04 $\pm$ 0.09 & 45 & 0.09 $\pm$ 0.02 & G & +0.02 $\pm$ 0.02 & 3 & +0.06 $\pm$ 0.03 & 4 & 4 \\
NGC 6494 & G & $-$0.04 $\pm$ 0.06 & 2 & 0.36     & T97 & $-$0.04 $\pm$ 0.08 & 3 & +0.03 $\pm$ 0.09 & 1 & 2 \\
NGC 6633 & G & $-$0.08 $\pm$ 0.08 & 5 & 0.20 $\pm$ 0.02 & G & $-$0.08 $\pm$ 0.12 & 4 & $-$0.04 $\pm$ 0.06 & 2 & 2 \\
NGC 6791 & G & +0.42 $\pm$ 0.08 & 2 & 0.15     & T97 & +0.42 $\pm$ 0.05 & 8 & +0.44 $\pm$ 0.05 & 4.5 & 5 \\
NGC 6811 & G & $-$0.01 $\pm$ 0.03 & 4 & 0.05     & [7]   & +0.03 $\pm$ 0.01 & 3 & $-$0.01    & 0.5 & 1 \\
NGC 6940 & G & +0.07 $\pm$ 0.06 & 11 & 0.20    & T97 &      +0.04 $\pm$ 0.02& 12   & +0.11 $\pm$ 0.06 & 2 & 2 \\
NGC 7092 & D & $-$0.02 $\pm$ 0.11 & 14 & 0.03 $\pm$ 0.02 & G &      &    & $-$0.01 $\pm$ 0.01 & 2.5 & 2 \\
NGC 7209 & G & $-$0.02 $\pm$ 0.01 & 2 & 0.15     & T97 &      &    & +0.03 $\pm$ 0.06 & 1 & 2 \\
NGC 7243 & D & +0.04 $\pm$ 0.12 & 8 & 0.24 $\pm$ 0.03 & G &      &    & +0.03 $\pm$ 0.01 & 2.5 & 2 \\
NGC 7789 & G & $-$0.02 $\pm$ 0.06 & 6 & 0.27     & T97 & +0.05 $\pm$ 0.07 & 5 & $-$0.05 $\pm$ 0.04 & 4 & 4 \\
\hline

\end{longtable}

\medskip
Notes and References: The new $Geneva$ metallicities, spectroscopic data by \citet{netopil16} and more recent works  - $a$ refers to results based on lower quality data, and updated weighted mean photometric metallicities with the total weight (W) and the number of different systems (No).
Calib: Type of calibration, dwarf stars (D) or giants (G); 
Ref: Reference of the reddening - (G) the MS reddening derived via $Geneva$ photometry is always given; (T97) \citep{twarog97}; (NP13) \citep{netopil13}; (ST) spectroscopic temperature and colour-temperature relation by \citet{huang15}; [1]: \citet{claria03}; [2]: \citet{claria96}; [3]: \citet{reddy15}; [4]: \citet{reddy13}; [5]: \citet{lapa00} downscaled; [6]: \citet{santos09}; [7]: \citet{molenda14}


\bsp	
\label{lastpage}
\end{document}